\documentclass{article}

\usepackage[english]{babel}
\usepackage{booktabs}
\usepackage{amssymb}
\usepackage{multirow}
\usepackage{tikz}
\usepackage{array}
\usepackage{parskip}
\usetikzlibrary{shapes.geometric, arrows.meta, positioning}

\usepackage{booktabs}
\usepackage{makecell}
\usepackage{graphicx}
\usepackage{parskip}
\usetikzlibrary{shapes.geometric, arrows.meta, positioning}
\usepackage{graphicx}
\usepackage{amsmath, amssymb, bm}
\usepackage{graphicx}
\usepackage{tikz}
\usepackage{ragged2e}
\usepackage{enumitem}

\usepackage[letterpaper,top=2cm,bottom=2cm,left=3cm,right=3cm,marginparwidth=1.75cm]{geometry}

\usepackage{amsmath}
\usepackage{graphicx}
\usepackage[colorlinks=true, allcolors=blue]{hyperref}

\title{Parsimonious Mixtures of Skewed Bilinear Factor Analyzers}
\author{Jacob Moore \& Michael Gallaugher}

\begin{document}
\maketitle
\begin{abstract}
Mixture models which cluster skewed random matrices can often suffer from over-parameterization in the absence of performing dimension reduction. Even with the use of bilinear factor analyzers, further parameter reduction can be achieved by constraining parameters over clusters. In this manuscript propose a parsimonious family of 256 models for mixtures of skewed matrix variate bilinear factor analyzers, specifically in the case of the skew $t$ distribution. An AECM algorithm for parameter estimation is discussed in detail. Further, extensive simulations are performed, and the method is considered in the case of the MNIST dataset and the Olivetti faces dataset.
\end{abstract}

\section{Introduction}

Three-way data is found in many modern applications where data on individuals sampled within a population are best represented as a matrix. Examples of three-way or matrix variate data are gray-scaled images as well as longitudinal multivariate data, where multiple variables are measured over time. If the population is heterogeneous, mixtures of matrix variate distributions can be used to model a sample. 

Classification is the process of categorizing individuals from a heterogeneous population. Clustering is a special case of classification where every observation in a sample is unlabeled, and the practice of mixture modeling has become commonplace in statistical classification and clustering. Classification procedures find and analyze underlying group structures in data. One common method used for clustering is model-based and generally makes use of a $G$-component finite mixture model. A multivariate random variable $\mathbf{X}$ from a finite mixture model has density:
\begin{align}
\label{eq:mix-mod-form}
f ( \mathbf{X} \mid \boldsymbol{\vartheta}) = \sum_{g = 1}^G \pi_g f_g ( \mathbf{X} \mid \boldsymbol{\theta}_g )
\end{align}
where $\boldsymbol{\vartheta} := (\pi_1, \ldots, \pi_G, \boldsymbol{\theta}_1, \ldots, \boldsymbol{\theta}_G)$, $f_g (\cdot)$ is the $g\textsuperscript{th}$ component density, $\boldsymbol{\theta}_g$ is the $g$th component's parameterization, and the mixing proportions, $\pi_1, \ldots, \pi_G$ are defined such that $\pi_g > 0$ for $g = 1,\ldots, G$ and $\sum_{g=1}^G \pi_g = 1$.

There has been much literature devoted to reducing dimensionality within the context of mixture models. \cite{banfield93}, \cite{celeux95}, and \cite{fraley02a} analyze mixtures of multivariate Gaussian distributions that share components within the Eigen decomposition of the covariance matrices. \cite{mcnicholas05} introduced the structure of the covariance matrix from factor analysis to mixture models and considered various models for the covariance matrices. These models are known as parsimonious Gaussian mixture models, and PGMMs can further reduce the number of parameters in these mixture models. 

The Gaussian distribution is a standard option for mixture modeling because of its mathematical tractability, but it lacks the flexibility to account for asymmetry or heavy tails. So, much work has been done in developing non-Gaussian mixture models. For example, \cite{peel00} and \cite{lin14} discuss mixtures of $t$-distributions. There's also discussion of asymmetric distributions such as the normal-inverse Gaussian distribution (\cite{karlis09}), the skew-$t$ distribution (\cite{lin10}, \cite{vrbik12}, etc.), the shifted asymmetric Laplace distribution (\cite{morris13b}), the variance-gamma distribution  (\cite{mcnicholas17}), and the generalized hyperbolic and its variations (\cite{browne15}, \cite{murray17b},\\ \cite{tang18}, \cite{tortora19}).

A naive approach to clustering matrix variate observations is to stack the columns of the matrix observations and treat the data as multivariate. Vectorization of these matrices and subsequently modeling the data in this form destroys the underlying matrix structure. Using matrix variate distributions also naturally reduce the number of free parameters required when estimating the covariance matrix. An early example of work in this area is \cite{viroli11} who considers mixtures of matrix variate normal distributions. Other work like \cite{gallaugher18a} introduces mixtures of skewed matrix variate distributions.

These approaches to mixture modeling three-way data still suffer from over-parameterization as the dimensionality of the observations increases. Literature in mixture modeling multivariate data has thoroughly addressed solutions for over-parameterization of the covariance matrices. Much of the work devoted to reducing parameters in multivariate settings have an analog in the matrix variate case. For example, \cite{gallaugher18b} discusses mixture models which make use of bilinear factor analyzers in the matrix variate normal case. This approach uses tools from bilinear probabilistic principal component analysis first discussed in \cite{zhao12} to reduce covariance parameters. This approach was also extended to skewed matrix variate distributions as well in \cite{gallaughermcnicholas2019}.

Another approach to reducing the number of free parameters in these matrix variate mixture models is to constrain parameters across groups. This concept is first explored by \\\cite{gallaughermcnicholas2020} which introduced parsimonious mixtures of matrix variate bilinear factor analyzers. This family is comprised of sixty-four parsimonious models of mixtures of Gaussian distributions. That is a large increase in models compared to the original 12 PGMMs from \cite{mcnicholas10d}. In the pursuit of reducing parameter counts for these mixture models, the computational costs of more complex modeling schemes has naturally grown.

Parsimonious mixtures of skewed matrix variate bilinear factor analyzers is a natural extension of the framework discussed in \cite{gallaughermcnicholas2020}. Mixture models of skewed matrix variate bilinear factor analyzers allow for flexible modeling of asymmetric or heavy-tailed data while also substantially reducing the number of parameters necessary for covariance estimation. We introduce the idea of parsimony to these models to further reduce potential over-parameterization. To combine this framework
with the parsimonious modeling discussed in \cite{gallaughermcnicholas2020} is not too challenging. However, the computational cost of fitting a large parsimonious family of skewed models could negate any practical benefits of this new methodology on its own, especially if constraints for higher order terms are considered. 

This paper introduces parsimonious mixtures of skewed bilinear factor analyzers which consists of families of two hundred fifty-six parsimonious models. The remainder of this work is laid out as follows: Some background on model-based clustering and matrix variate methods are presented in Section 2. The methodology is outlined in Section 3. Important numerical considerations are outlined in Section 4. Then, simulations and data analyses are presented in Sections 5, and we conclude with a discussion of these results in Section 7.

\section{Background}

\subsection{Generalized inverse Gaussian distribution}

A random variable $Y$ has a generalized inverse Gaussian (GIG) distribution, with parameters $a$, $b$, and $\lambda$, denoted by $\text{GIG}(a, b, \lambda)$, if it's probability density function can be written in the form:
$$
f \left( y \mid a, b, \lambda\right) = \frac{(a / b)^{\lambda / 2} y^{\lambda - 1}}{2 K_{\lambda} (\sqrt{ab})} \exp \left\{ - \frac{ay + b / y}{2}\right\}.
$$
For $y > 0$, $a, b \in \mathbb{R}^+$, and $\lambda \in \mathbb{R}$, where 
\begin{equation}
\label{eq:bessel-def}
K_{\lambda}(u) = \frac{1}{2} \int_{0}^{\infty} y^{\lambda -1} \exp \left[ -\frac{u}{2} \left( y + \frac{1}{y}\right)\right]\, dy
\end{equation}
is the modified Bessel function of the second kind with index $\lambda$. Expectations for some functions of a GIG random variable are
mathematically tractable in terms of $K_{\lambda} (u)$, e.g.:
\begin{equation}
\begin{aligned}
\label{eq:gig-exp}
\mathbb{E} [Y] = \sqrt{\frac{b}{a}} \frac{K_{\lambda + 1} (\sqrt{ab})}{K_{\lambda} (\sqrt{ab})}, &\quad
\mathbb{E} \left[ 1/ Y\right] = \sqrt{\frac{a}{b}} \frac{K_{\lambda + 1} (\sqrt{ab})}{K_{\lambda} (\sqrt{ab})} - \frac{2\lambda}{b}, \\
\mathbb{E} \left[ \log Y\right] = \log \left( \sqrt{\frac{b}{a}}\right) &+ \frac{1}{K_{\lambda} (\sqrt{ab})} \frac{d}{d \xi}  K_{\xi} (\sqrt{ab}) \bigg|_{\xi = \lambda}.
\end{aligned}
\end{equation}

\subsection{Matrix variate distributions}

Modeling three-way data can be facilitated through the use of matrix variate distributions. The most tractable and widely known matrix variate distribution is the matrix variate normal distribution. Let $\mathbf{X}$ denote a realization of the random matrix $\mathcal{X}$. Suppose $\mathcal{X}$ is an $n \times p$ random matrix with a matrix variate normal distribution with an $n \times p$ location parameter $\mathbf{M}$ and positive definite scale parameters $\boldsymbol{\Sigma}$ and $\boldsymbol{\Psi}$ with dimension $n \times n$ and $p \times p$, respectively. Then, we say $\mathcal{X} \sim \mathcal{N}_{n \times p} \left(\mathbf{M}, \boldsymbol{\Sigma}, \boldsymbol{\Psi}\right)$, and the probability density function of $\mathcal{X}$ can be expressed as:
$$
\phi_{n \times p} \left( \mathbf{X} \mid \mathbf{M}, \boldsymbol{\Sigma}, \boldsymbol{\Psi}\right) =
\frac{1}{(2 \pi)^{\frac{np}{2}} |\boldsymbol{\Sigma}|^{\frac{p}{2}} |\boldsymbol{\Psi}|^{\frac{n}{2}}} \exp \left\{ -\frac{1}{2} \boldsymbol{\Sigma}^{-1} \left(\mathbf{X} - \mathbf{M}\right) \boldsymbol{\Psi}^{-1} \left(\mathbf{X} - \mathbf{M}\right)^{\prime}\right\}.
$$

Shown in \cite{harrar08}, the matrix variate normal distribution has a direct relationship with the multivariate normal distribution. In other words, $\mathcal{X} \sim \mathcal{N}_{n \times p} \left(\mathbf{M}, \boldsymbol{\Sigma}, \boldsymbol{\Psi}\right)$ if and only if $\operatorname{vec} \left( \mathcal{X}\right) \sim \mathcal{N}_{np} \left(\operatorname{vec}\left(\mathbf{M}\right), \boldsymbol{\Psi} \otimes \boldsymbol{\Sigma} \right)$ where $\operatorname{vec}\left(\cdot\right)$ is vectorization operator, $\mathcal{N}_{np}(\cdot)$ is the multivariate normal distribution of dimension $n \times p$, and $\otimes$ is the Kronecker product. \cite{harrar08} discusses the topic of matrix variate skew-normal distributions, and there are other matrix variate distributions discussed in statistical literature such as the well known Wishart distribution. In this paper, the models will utilize matrix variate distributions first introduced by \cite{gallaugher17a} and \cite{gallaugher17b}. These skewed matrix variate distributions will function as the component distributions of the mixture models. These distributions assume that a random matrix $\mathcal{X}$ can be expressed in the following form:
\begin{align}
\label{eq:mv-form}
\mathcal{X} = \mathbf{M} + W \mathbf{A} + \sqrt{W} \mathcal{V}
\end{align}
where $\mathbf{M}$ and $\mathbf{A}$ are both $n \times p$ matrices which represent location and skewness, respectively. $\mathcal{V} \sim \mathcal{N}_{n \times p} \left(\boldsymbol{0}, \boldsymbol{\Sigma}, \boldsymbol{\Psi}\right)$, and $W > 0$ is a random variable with some distribution with a probability density function: $h \left(w \mid \boldsymbol{\theta}\right)$.

In \cite{gallaugher17a}, the matrix variate skew-$t$ distribution is defined as a special case of \eqref{eq:mv-form} where $W$ has an inverse gamma distribution $W \sim \mathrm{IGamma}(a, b)$. The probability density function of $W$ is expressed as:
\begin{equation}
h \left(w \mid a, b\right) = \frac{b^a}{\Gamma (a)} w^{-a-1} \exp \left\{ -\frac{b}{w}\right\},
\end{equation}
where $\Gamma (\cdot)$ denotes the Gamma function. If $W \sim \mathrm{IGamma}(\nu/2, \nu/2)$, the resulting density of $\mathcal{X}$ can be expressed as:
\begin{equation}
\begin{aligned}
\label{eq:st-pdf}
f_{\mathrm{MVST}}(\mathbf{X}\mid\boldsymbol{\vartheta})
&=
\frac{
2\left(\frac{\nu}{2}\right)^{\nu/2}
\exp\!\left\{\operatorname{tr}\!\big(\boldsymbol{\Sigma}^{-1}(\mathbf{X}-\mathbf{M})\boldsymbol{\Psi}^{-1}\mathbf{A}^{\prime}\big)\right\}
}{
(2\pi)^{np/2}\,|\boldsymbol{\Sigma}|^{p/2}\,|\boldsymbol{\Psi}|^{n/2}\,\Gamma\!\left(\frac{\nu}{2}\right)
}
\left(
\frac{\delta(\mathbf{X};\mathbf{M},\boldsymbol{\Sigma},\boldsymbol{\Psi})+\nu}{\rho(\mathbf{A},\boldsymbol{\Sigma},\boldsymbol{\Psi})}
\right)^{-(\nu+np)/4}
\\ &\qquad\times
\,K_{-(\nu+np)/2}\!\left(
\sqrt{\left|\rho(\mathbf{A},\boldsymbol{\Sigma},\boldsymbol{\Psi})\right|
\left|\delta(\mathbf{X};\mathbf{M},\boldsymbol{\Sigma},\boldsymbol{\Psi})+\nu\right|}
\right).    
\end{aligned}
\end{equation}
where
\begin{equation}
\begin{aligned}
\label{eq:delta-rho}
\delta(\mathbf{X}; \mathbf{M}, \boldsymbol{\Sigma}, \boldsymbol{\Psi})
&:=
\operatorname{tr}\!\left(
\boldsymbol{\Sigma}^{-1}(\mathbf{X}-\mathbf{M})
\boldsymbol{\Psi}^{-1}(\mathbf{X}-\mathbf{M})'
\right),    \\
\rho(\mathbf{A}; \boldsymbol{\Sigma}, \boldsymbol{\Psi})
&:=
\operatorname{tr}\!\left(
\boldsymbol{\Sigma}^{-1}\mathbf{A}
\boldsymbol{\Psi}^{-1}\mathbf{A}'
\right),
\end{aligned}
\end{equation}
and $\nu > 0$. In summary, the following notation: $\mathcal{X} \sim\mathrm{MVST}_{n \times p}(\mathbf{M}, \mathbf{A}, \boldsymbol{\Sigma}, \boldsymbol{\Psi}, \nu)$ is how a matrix variate skew-t random variable will be referenced through this paper.

\subsection{Mixtures of Matrix Variate Bilinear Factor Analyzers}

Consider a mixture of skewed bilinear factor analyzers corresponding to one of the skewed distributions mentioned above. Each of these distributions is derived from a matrix variate normal variance-mean mixture. That is, each random matrix $\mathcal{X}_i$ can be written as:
\begin{align}
    \mathcal{X}_i = \mathbf{M}_g + W_{ig} \mathbf{A}_g + \mathcal{V}_{ig}
\end{align}
with probability $\pi_g$ for $g \in \{1, 2, \ldots, G\}$. Here, $\mathbf{M}_g$ is the location of the $g$th component, $\mathbf{A}_g$ is the skewness, and $W_{ig}$ is a random variable with a density $h(w_{ig} | \boldsymbol{\theta}_g)$ which controls for the kurtosis in the matrix variate distribution. We also assume that $\mathcal{V}_{ig}$ can be written as:
\begin{align}
    \mathcal{V}_{ig} = \boldsymbol{\Lambda}_g \mathcal{U}_{ig} \boldsymbol{\Delta}_{g}^{\prime} + \boldsymbol{\Lambda}_g \mathcal{E}^{B}_{ig} + \mathcal{E}^{A}_{ig} \boldsymbol{\Delta}_{g}^{\prime} + \mathcal{E}_{ig}
\end{align}
where $\boldsymbol{\Lambda}_g$ is a $n \times q$ matrix of column factor loadings, with $q < n$. $\boldsymbol{\Delta}_{g}$ is a $p \times r$ matrix of row factor loadings, with $r < p$. And the random matrices have the following conditional distributions:
\begin{align}
    &\mathcal{U}_{ig} \mid w_{ig} \sim \mathcal{N}_{q \times r} (\boldsymbol{0}, w_{ig} \boldsymbol{I}_q, \boldsymbol{I}_r),
    &\mathcal{E}^{B}_{ig} \mid w_{ig} \sim \mathcal{N}_{q \times p} (\boldsymbol{0}, w_{ig} \boldsymbol{I}_q, \boldsymbol{\Psi}_g), \\
    &\mathcal{E}^{A}_{ig} \mid w_{ig} \sim \mathcal{N}_{n \times r} (\boldsymbol{0}, w_{ig} \boldsymbol{\Sigma}_g, \boldsymbol{I}_r),
    &\mathcal{E}_{ig} \mid w_{ig} \sim \mathcal{N}_{n \times p} (\boldsymbol{0}, w_{ig} \boldsymbol{\Sigma}_g, \boldsymbol{\Psi}_g).
\end{align}

It is assumed that $\mathcal{U}_{ig}, \mathcal{E}^{B}_{ig}, \mathcal{E}^{A}_{ig}$ and $\mathcal{E}_{ig}$ are mutually independent. Additionally, the scale matrices $\boldsymbol{\Sigma}_g$ and $\boldsymbol{\Psi}_g$ are diagonal and positive definite.

To facilitate clustering, the standard approach is to introduce the indicator $Z_{ig}$ which is 1 if the $i$th observation belongs to the $g$th component and 0 otherwise. Then, it can be shown that
\begin{align}
    \mathcal{X}_i \mid Z_{ig} = 1 \sim \mathrm{MVST}_{n \times p} (\mathbf{M}_g, \mathbf{A}_g, \boldsymbol{\Sigma}_g + \boldsymbol{\Lambda}_g \boldsymbol{\Lambda}_g^{\prime}, \boldsymbol{\Psi}_g + \boldsymbol{\Delta}_{g} \boldsymbol{\Delta}_{g}^{\prime}, \nu_g)
\end{align}

Where $\boldsymbol{D}_{n \times p}$ represents one of the four matrix variate distributions discussed in the Matrix Variate Distributions subsection with distribution-specific parameters $\boldsymbol{\theta}_g$.
Following previous methods like what's done in \cite{zhao12} and \cite{gallaughermcnicholas2019}, this model has a two-stage interpretation given by
$$
\begin{aligned}
\mathcal{X}_i 
&= \mathbf{M}_g + \mathbf{W}_{ig}\mathbf{A}_g 
   + \boldsymbol{\Lambda}_g \boldsymbol{\mathcal{Y}}^{B}_{ig} 
   + \mathcal{R}^{B}_{ig}, \\[6pt]
\boldsymbol{\mathcal{Y}}^{B}_{ig}
&:= \boldsymbol{\mathcal{U}}_{ig} \boldsymbol{\Delta}_g^{\prime}
   + \mathcal{E}^{B}_{ig}, \\[6pt]
\mathcal{R}^{B}_{ig}
&:= \mathcal{E}^{A}_{ig} \boldsymbol{\Delta}_g^{\prime}
   + \mathcal{E}_{ig},
\end{aligned}
$$
and
$$
\begin{aligned}
\mathcal{X}_i 
&= \mathbf{M}_g + \mathbf{W}_{ig}\mathbf{A}_g 
   + \boldsymbol{\mathcal{Y}}^{A}_{ig} \boldsymbol{\Delta}_g^{\prime}
   + \mathcal{R}^{A}_{ig}, \\[6pt]
\boldsymbol{\mathcal{Y}}^{A}_{ig}
&:= \boldsymbol{\Lambda}_g \boldsymbol{\mathcal{U}}_{ig}
   + \mathcal{E}^{A}_{ig}, \\[6pt]
\mathcal{R}^{A}_{ig}
&:= \boldsymbol{\Lambda}_g \mathcal{E}^{B}_{ig}
   + \mathcal{E}_{ig}.
\end{aligned}
$$
These forms will be useful in performing parameter estimation of the scale matrices.

\subsection{Parsimonious MMVBFA models} 

\cite{gallaughermcnicholas2020} discusses parsimonious mixtures of matrix variate bilinear factor analyzers, a matrix variate extension of the better known parsimonious Gaussian mixture models first first discussed in \cite{mcnicholas05}, designed for multivariate data. PMMVBFA models, as their referred to, total to 64 different models using mixtures of matrix variate normal distributions. Combinations of the following constraints: $\boldsymbol{\Lambda}_g = \boldsymbol{\Lambda}$, $\boldsymbol{\Sigma}_g = \boldsymbol{\Sigma}$, $\boldsymbol{\Sigma}_g = \sigma_g \boldsymbol{I}_n$ ($\sigma_g \in \mathbb{R}^{+}$), $\boldsymbol{\Delta}_g = \boldsymbol{\Delta}$, $\boldsymbol{\Psi}_g = \boldsymbol{\Psi}$, $\boldsymbol{\Psi}_g = \psi_g \boldsymbol{I}_p$ ($\psi_g \in \mathbb{R}^{+}$) are considered, and maximum likelihood estimation is performed in a similar way as it is here. Table \ref{tab:rowmodels} and Table \ref{tab:colmodels} come directly from \cite{gallaughermcnicholas2020} and give the number of free parameters contributed by the scale covariance matrices for a given row model or column model. These same configurations of scale matrices will also be considered for the skewed matrix variate distributions discussed in this paper.

\begin{table}
\label{tab:rowmodels}
\caption{Row Models and their respective number of contributed parameters}
$$
\begin{array}{cccc}
\hline
\boldsymbol{\Lambda}_g = \boldsymbol{\Lambda} & \boldsymbol{\Sigma}_g = \boldsymbol{\Sigma} & \boldsymbol{\Sigma}_g = \sigma_g \mathbf{I}_n & \text{Number of Scale Parameters} \\
\hline
\text{C} & \text{C} & \text{C} & [nq - q(q - 1)/2] + 1 \\
\text{C} & \text{C} & \text{U} & [nq - q(q - 1)/2] + n \\
\text{C} & \text{U} & \text{C} & [nq - q(q - 1)/2] + G \\
\text{C} & \text{U} & \text{U} & [nq - q(q - 1)/2] + nG \\
\text{U} & \text{C} & \text{C} & G[nq - q(q - 1)/2] + 1 \\
\text{U} & \text{C} & \text{U} & G[nq - q(q - 1)/2] + n \\
\text{U} & \text{U} & \text{C} & G[nq - q(q - 1)/2] + G \\
\text{U} & \text{U} & \text{U} & G[nq - q(q - 1)/2] + nG \\
\hline
\end{array}
$$
\end{table}

\begin{table}
\label{tab:colmodels}
\caption{Column Models and their respective number of contributed parameters}
$$
\begin{array}{cccc}
\hline
\boldsymbol{\Delta}_g = \boldsymbol{\Delta} & \boldsymbol{\Psi}_g = \boldsymbol{\Psi} & \boldsymbol{\Psi}_g = \psi_g \mathbf{I}_r & \text{Number of Scale Parameters} \\
\hline
\text{C} & \text{C} & \text{C} & [pr - r(r - 1)/2] + 1 \\
\text{C} & \text{C} & \text{U} & [pr- r(r - 1)/2] + p \\
\text{C} & \text{U} & \text{C} & [pr - r(r - 1)/2] + G \\
\text{C} & \text{U} & \text{U} & [pr - r(r - 1)/2] + pG \\
\text{U} & \text{C} & \text{C} & G[pr - r(r - 1)/2] + 1 \\
\text{U} & \text{C} & \text{U} & G[pr - r(r - 1)/2] + p \\
\text{U} & \text{U} & \text{C} & G[pr - r(r - 1)/2] + G \\
\text{U} & \text{U} & \text{U} & G[pr - r(r - 1)/2] + pG \\
\hline
\end{array}
$$
\end{table}

\section{Methodology}

\subsection{Parsimonious mixtures of skewed bilinear factor analyzers}

As discussed in \cite{gallaughermcnicholas2020}, MMVBFA models lend themselves naturally to a matrix variate extension of the PGMM models. For these models, these combinations of the constraints are considered: $\boldsymbol{\Lambda}_g = \boldsymbol{\Lambda}$, $\boldsymbol{\Sigma}_g = \boldsymbol{\Sigma}$, $\boldsymbol{\Sigma}_g = \sigma_g \boldsymbol{I}_n$ ($\sigma_g \in \mathbb{R}^{+}$), $\boldsymbol{\Delta}_g = \boldsymbol{\Delta}$, $\boldsymbol{\Psi}_g = \boldsymbol{\Psi}$, and $\boldsymbol{\Psi}_g = \psi_g \boldsymbol{I}_p$ ($\psi_g \in \mathbb{R}^{+}$). These constraints are the same PGMM constraints applied to both scale covariance matrices. One could also consider applying extended PGMM constraints the covariance matrices similar to \cite{mcnicholas10d}, and this is still a possible extension of the PMMVBFA models. However, models for mixtures of skewed matrix variate distributions come at the cost of also having to estimate $\mathbf{A}_g$ and $\nu_g$. So, instead of further reducing free parameters on the scale matrices, this paper considers possible constraints on parameters controlling for higher order moments. We propose the constraints $\mathbf{A}_g = \mathbf{A}$ and $\nu_g = \nu$. With eight different constraints, there are 256 different constrained models in total which will fit every combination of these. These models are referred to as the Parsimonious Mixtures of Skewed Bilinear Factor Analyzers (PMSBFA) family. In Tables \ref{tab:rowmodels} and \ref{tab:colmodels}, parts of the models along with the number of scale parameters are specified for the row and column scale matrices. These are referred to as row models and column models respectively. Additionally, each skewness matrix $\mathbf{A}_g$ adds an $np$ free parameters, totaling to $Gnp$ free parameters. So, constraining $\mathbf{A}_g$ across groups would mean there are only $np$ free parameters need to be estimated for the shared skewness matrix. Lastly, the number of concentration parameters is $G$ if left unconstrained and only 1 if we impose $\nu_g = \nu$.

Parameter estimates that are fit with a subset of these outlined constraints are often referred to directly by the nomenclature used in this paper. These model fits are referred to by their ``constraint model". A constraint model is a string of four blocked labels that are spaced using hyphens, and the blocked labels in a constrained model are ordered to represent the constraints on the row model, the column model, the skewness constraint, and its concentration. For example, CUC-UCU-U-C refers to a PMSBFA model with a row model CUC ($\boldsymbol{\Lambda}_g = \boldsymbol{\Lambda}$, $\boldsymbol{\Sigma}_g = \sigma_g \boldsymbol{I}_n$), column model UCU ($\boldsymbol{\Delta}_g = \boldsymbol{\Delta}_g$, $\boldsymbol{\Psi}_g = \boldsymbol{\Psi}$), unconstrained skewness ($\mathbf{A}_g = \mathbf{A}_g$), and a constrained concentration parameter ($\nu_g = \nu$).

Maximum likelihood estimation is performed using an alternating expectation maximization (AECM) algorithm in a nearly identical fashion to \cite{gallaughermcnicholas2019}. However, the maximization steps will differ depending on the constrained model which is being fit. Introduced below are the three stages of the AECM algorithm and the various maximization steps which can be broken down by row and column model.

\subsection{Parameter estimation with the AECM algorithm}

Suppose a random sample of size $N$ of $n \times p$ random matrices, denoted by $\mathbf{X} := (\mathbf{X}_1,\ldots,\mathbf{X}_N$), are observed. The observed log-likelihood at the $t$th iteration is defined as:

\begin{equation}
\begin{aligned}
\label{eq:oll}
\ell \left( \hat{\boldsymbol{\vartheta}}^{(t)} \mid \mathbf{X}\right) := \sum_{i=1}^{N} \log \left( \sum_{g=1}^G \hat{\pi}^{(t)}_g f \left(\mathbf{X}_i \mid \hat{\boldsymbol{\vartheta}}_{g}^{(t)}\right)\right),
\end{aligned}
\end{equation}

where $f$ is the probability density function of $\mathcal{X} \mid Z_{ig} = 1$, $\hat{\boldsymbol{\vartheta}}^{(t)}_g$ represents the set of parameter estimates associated with the $g\textsuperscript{th}$ component at iteration $t$, and $\hat{\boldsymbol{\vartheta}}^{(t)} := \left(\hat{\pi}^{(t)}_1, \ldots, \hat{\pi}^{(t)}_G, \hat{\boldsymbol{\theta}}^{(t)}_1, \ldots, \hat{\boldsymbol{\theta}}^{(t)}_G\right)$. Later sections of this paper may also refer to the observed log-likelihood observed on a finalized set of model parameters ($\hat{\boldsymbol{\vartheta}}$) as $\ell (\hat{\boldsymbol{\vartheta}})$.

\subsection*{AECM stage 1}

The complete log-likelihood can be expressed as follows with the inclusion of the component memberships $\boldsymbol{Z} : = (\boldsymbol{Z}_1, \ldots, \boldsymbol{Z}_N)$ and the latent random vectors $\boldsymbol{W} : = (\boldsymbol{W}_1,\ldots, \boldsymbol{W}_N)$ where $\boldsymbol{Z}_i= (z_{i1}, \ldots, Z_{ig})^\prime$ and $\boldsymbol{W}_i = (W_{i1}, W_{i2}, \ldots, W_{iG})^\prime$ for $i= 1, \ldots, N$:
\begin{equation}
\begin{aligned}
\label{eq:cll-1}
\ell_{\text{C}_1} &= \mathcal{C} + \sum_{i=1}^N \sum_{g=1}^G Z_{ig} \bigg[
\log \pi_g + \log h(w_{ig} \mid \boldsymbol{\theta}_g)
- \frac{1}{2} \operatorname{tr} \bigg\{
\frac{1}{W_{ig}} \left(\boldsymbol{\Sigma}_g^*\right)^{-1} \left(\mathbf{X}_i - \mathbf{M}_g\right) \left(\boldsymbol{\Psi}_g^*\right)^{-1} \left(\mathbf{X}_i - \mathbf{M}_g\right)'  \\
&\qquad\qquad - 2\left(\boldsymbol{\Sigma}_g^*\right)^{-1} \left(\mathbf{X}_i - \mathbf{M}_g\right)\left(\boldsymbol{\Psi}_g^*\right)^{-1} \mathbf{A}_g'
+ W_{ig} \left(\boldsymbol{\Sigma}_g^*\right)^{-1} \mathbf{A}_g \left(\boldsymbol{\Psi}_g^*\right)^{-1} \mathbf{A}_g'
 \bigg\} \bigg],
\end{aligned}
\end{equation}

where $\mathcal{C}$ is a constant,
$\boldsymbol{\Sigma}_g^* := \boldsymbol{\Lambda}_g \boldsymbol{\Lambda}_g' + \boldsymbol{\Sigma}_g$, and $\boldsymbol{\Psi}_g^* := \boldsymbol{\Delta}_g \boldsymbol{\Delta}_g' + \boldsymbol{\Psi}_g$. In the E-step, for the $i$th observation, we calculate the following conditional expectation for the $g$th component assignment as:
\begin{equation}
\label{eq:z-update}
\hat{Z}_{ig}^{(t+1)} = \frac{\hat{\pi}^{(t)}_g f\left(\mathbf{X}_i \mid \hat{\boldsymbol{\vartheta}}_g^{(t)}\right)}{
\sum_{h=1}^G \hat{\pi}^{(t)}_h f\left(\mathbf{X}_i \mid \hat{\boldsymbol{\vartheta}}_h^{(t)}\right)}
\end{equation}
where $f$ represents the $g\textsuperscript{th}$ component density evaluated for $\mathbf{X}_i$. Functions of the latent variable $W_{ig}$ appear in $\ell_{\text{C}_1}$. So, the following symbols are defined  used to represent those conditional expectations: 
\begin{equation}
\label{eq:cond-exp-def}
a_{ig}^{(t+1)}\!\! :=\! \mathbb{E} \!\bigg[ \!W_{ig} \bigg|\mathbf{X}_i, Z_{ig} \!=\! 1, \hat{\boldsymbol{\vartheta}}_g^{(t)} \!\bigg]\!, \,\,
b_{ig}^{(t+1)} \!\!:= \!\mathbb{E}\! \bigg[ \! \frac{1}{W_{ig}}  \bigg| \mathbf{X}_i, Z_{ig} \!=\! 1, \hat{\boldsymbol{\vartheta}}_g^{(t)} \!\bigg]\!, \,\,
c_{ig}^{(t+1)} \!\!:=\! \mathbb{E} \!\bigg[ \!\log W_{ig} \bigg|\mathbf{X}_i, Z_{ig} \!=\! 1, \hat{\boldsymbol{\vartheta}}_g^{(t)}\!\bigg]\!.
\end{equation}
The conditional expectations here can be calculated using \eqref{eq:gig-exp}. So, $a_{ig}^{(t+1)}$, $b_{ig}^{(t+1)}$, and $c_{ig}^{(t+1)}$ have a closed form in terms of modified Bessel functions of the second kind since the conditional distribution of the latent variable $W_{ig}$, given $\mathbf{X}_i$ and component membership $Z_{ig} = 1$, was shown in \cite{gallaugher17a} and \cite{gallaugher17b} to have a generalized inverse Gaussian distribution with specific parameters depending on the initial distribution of $W_{ig}$. For the skew $t$ distribution, the conditional distribution of $W_{ig}$ is as follows:
$$
\begin{aligned}
W_{ig}^{\mathrm{ST}} \mid \mathbf{X}_i,\, Z_{ig}=1
&\sim \operatorname{GIG}\!\left(
\rho(\mathbf{A}_g, \boldsymbol{\Sigma}_g^*, \boldsymbol{\Psi}_g^*),
\delta(\mathbf{X}_i; \mathbf{M}_g, \boldsymbol{\Sigma}_g^*, \boldsymbol{\Psi}_g^*) + \nu_g,
-\frac{\nu_g + np}{2}
\right).
\end{aligned}
$$

\subsection*{Stage 1 updates}

The M-step for some parameters in this stage will depend on the skewness constraint and the concentration constraint. However, regardless of the constraints, the updates for the mixing proportions are: 
$$
\hat{\pi}^{(t + 1)}_g = \frac{N_g}{N}, \text{ where } N_g := \sum_{i=1}^N \hat{z}^{(t + 1)}_{ig}.
$$
This can be shown with the use of Lagrange multipliers on the expectation of \eqref{eq:cll-1}. The main parameters that are updated in this first maximization stage are the location and skewness parameters. For a model with unconstrained skewness, we update $\hat{\mathbf{M}}_g$ and $\hat{\mathbf{A}}_g$ for $g = 1, \dots, G$ in exact same fashion as what's done in \cite{gallaugher18a} and \\
\cite{gallaughermcnicholas2019}. The updates are as follows:
\begin{align}
\label{eq:orig-ms1-updates}
\hat{\mathbf{M}}^{(t + 1)}_g = \frac{ \sum_{i=1}^N \hat{z}^{(t + 1)}_{ig} \left( \bar{a}_g b^{(t + 1)}_{ig} - 1 \right) \mathbf{X}_i }{ N_g \left(\bar{a}_g \bar{b}_{g} - 1\right) }, \quad
\hat{\mathbf{A}}^{(t + 1)}_g = \frac{ \sum_{i=1}^N \hat{z}^{(t + 1)}_{ig} \left( \bar{b}_g - b^{(t + 1)}_{ig} \right) \mathbf{X}_i }{ N_g \left(\bar{a}_g \bar{b}_{g} - 1\right) },
\end{align}
where
$$
\bar{a}_g := \frac{1}{N_g} \sum_{i=1}^N \hat{z}^{(t + 1)}_{ig} a^{(t + 1)}_{ig}, \quad
\bar{b}_g := \frac{1}{N_g} \sum_{i=1}^N \hat{z}^{(t + 1)}_{ig} b^{(t + 1)}_{ig}.
$$
These updates come from finding values for which the gradient of the expectation of \eqref{eq:cll-1} is zero. With no constraints on $\mathbf{A}_g$, this amounts to solving this system: 
$$
\frac{\partial}{\partial \mathbf{M}_g} \mathbb{E}_{\boldsymbol{Z}, \boldsymbol{W}} \left[ \ell_{\text{C}_1} \mid \mathbf{X}, \hat{\boldsymbol{\vartheta}}^{(t)}\right] = \boldsymbol{0}, \qquad
\frac{\partial}{\partial \mathbf{A}_g} \mathbb{E}_{\boldsymbol{Z}, \boldsymbol{W}} \left[ \ell_{\text{C}_1} \mid \mathbf{X}, \hat{\boldsymbol{\vartheta}}^{(t)}\right] = \boldsymbol{0}, 
$$ 
for each $g = 1, \ldots, G$, where $\mathbb{E}_{\boldsymbol{Z}, \boldsymbol{W}} \left[ \cdot \mid \mathbf{X}, \hat{\boldsymbol{\vartheta}}^{(t)}\right]$ is the expectation operator with respect to the unknown latent variables $\boldsymbol{Z}$ and $\boldsymbol{W}$, conditional on $\mathbf{X}$ and $\hat{\boldsymbol{\vartheta}}^{(t)}$. This approach results $G$ linear systems of equations with two $n \times p$ unknown matrices with closed form solutions \eqref{eq:orig-ms1-updates}. However, with the restriction $\mathbf{A}_g = \mathbf{A}$ in place, equating the gradient to zero no longer results in $G$ separate linear systems. Instead, we first update $\hat{\mathbf{A}}$ as the matrix that satisfies this linear system:
\begin{equation}
\begin{aligned}
\label{eq:const-skew-upd}
\sum_{g = 1}^G \!N_g \!\left( \bar{a}_g \!- \!\frac{1}{\bar{b}_g} \right) \!\!\left[ \left( \!\hat{\boldsymbol{\Psi}}_{\!g}^*\!\right)^{\!\!-1} \!\!\!\otimes\! \left(\!\hat{\boldsymbol{\Sigma}}_{g}^*\!\right)^{\!\!-1}\right]\!\operatorname{vec}\! \left( \hat{\mathbf{A}}^{\!(t + 1)}\right)\! = \! \operatorname{vec} \!\left( \sum_{g = 1}^G \!\left(\!\hat{\boldsymbol{\Sigma}}_g^*\!\right)^{\!\!-1}\!\! \left[
\sum_{i = 1}^N \hat{Z}_{ig}^{(t + 1)} \!\!\left( \!1 \!-\! \frac{b_{ig}^{(t + 1)}}{\bar{b}_g}\!\right) \!\mathbf{X}_i
\right] \!\left(\!\hat{\boldsymbol{\Psi}}_{\!g}^*\!\right)^{\!\!-1}\!\right)\!,
\end{aligned}
\end{equation}
where $\hat{\boldsymbol{\Sigma}}_g^* := \hat{\boldsymbol{\Sigma}}_g^{(t)} + \hat{\boldsymbol{\Lambda}}_g^{(t)}\hat{\boldsymbol{\Lambda}}_g^{\prime(t)}$ and $\hat{\boldsymbol{\Psi}}_g^* := \hat{\boldsymbol{\Psi}}_g^{(t)} + \hat{\boldsymbol{\Delta}}_g^{(t)}\hat{\boldsymbol{\Delta}}_g^{\!\prime(t)}$. This calculation of $\hat{\mathbf{A}}$ involves solving a linear system of equations with $np$ unknowns. For the sake of computational efficiency, we take advantage of the fact that the $np \times np$ matrix in \eqref{eq:const-skew-upd} is symmetric and positive definite, and numerical evaluation of the left hand side does not require the full computational cost of an $np \times np$ matrix multiplication. This is the ideal scenario for iterative methods such as the Conjugate Gradient method. After the shared skewness matrix is updated, each location matrix is updated as:
\begin{equation}
\hat{\mathbf{M}}_g^{(t + 1)} = \frac{1}{N_g\bar{b}_g} \left[\sum_{i = 1}^N \hat{z}^{(t + 1)}_{ig} b^{(t + 1)}_{ig} \mathbf{X}_i\right] - \frac{1}{\bar{b}_g} \hat{\mathbf{A}}^{(t + 1)}
\end{equation}
The values of $\hat{\nu}_g^{(t+1)}$ are also updated in this M-step. The part of the complete log-likelihood affected by the degrees of freedom can be expressed as:
$$
\begin{aligned}
\mathcal{L}^{\mathrm{MVST}} = \sum_{i=1}^N \sum_{g =1}^G Z_{ig} \log h \left(W_{ig} \mid \nu_g\right) =
\sum_{i=1}^{N}\sum_{g=1}^{G} Z_{ig}\!
\left[
\frac{\nu_g}{2}\log\!\left(\frac{\nu_g}{2}\right)
\!-\!
\log\!\left(\Gamma\!\left(\frac{\nu_g}{2}\right)\right)
\!-\!
\frac{\nu_g}{2}\left(
\!\log(w_{ig}) \!+\! \frac{1}{w_{ig}}
\right)
\!\right].
\end{aligned}
$$
For the unconstrained system ($\nu_g = \nu_g$), the update $\hat{\nu}_g^{(t + 1)}$ is the value of $\nu_g$ that satisfies the following equation, the same equation as the one shown in \cite{gallaugher18a}:
$$
\begin{aligned}
\log\!\left(\frac{\nu_g}{2}\right) + 1 - \varphi\!\left(\frac{\nu_g}{2}\right) 
- (\bar{b}_g + \bar{c}_g) = 0,
\end{aligned}
$$
where $\varphi\!(\cdot)$ denotes the digamma function and $\bar{c}_g = N_g^{-1} \sum_{i = 1}^n \hat{z}^{(t+1)}_{ig} c_{ig}^{(t+1)}$. For the a model with constrained degrees of freedom across components ($\nu_g = \nu$), the update $\hat{\nu}^{(t + 1)}$ is the value of $\nu$ that satisfies the
following equation:
$$
\begin{aligned}
\log\!\left(\frac{\nu}{2}\right) + 1 - \varphi\!\left(\frac{\nu}{2}\right) 
- \frac{1}{N} \sum_{i=1}^{N} \sum_{g = 1}^G \hat{z}^{(t + 1)}_{ig} \left( b^{(t + 1)}_{ig} + c^{(t + 1)}_{ig} \right) = 0.
\end{aligned}
$$

\subsection*{AECM stage 2}

For the calculation of $\boldsymbol{\Sigma}_g$ and $\boldsymbol{\Lambda}_g$ for $g = 1,\ldots, G$, we consider the complete log-likelihood with the same information as in stage 1 with the addition of $\boldsymbol{\mathcal{Y}}_{ig}^B$ for $g = 1,\ldots, G$ for each $i = 1\,\ldots,N$. This complete log-likelihood can be expressed as:
$$
\begin{aligned}
\ell_{\mathrm{C}_2} &= \mathcal{C} + \sum_{i=1}^N \sum_{g=1}^G Z_{ig} \left[
\log \pi_g + \log h(W_{ig} \mid \boldsymbol{\theta}_g) + \log \phi_{q \times p}(\boldsymbol{\mathcal{Y}}_{ig}^B \mid \boldsymbol{0}, W_{ig} \mathbf{I}_q, \boldsymbol{\Psi}_g^*)\right. \\
&\quad \left. 
+ \log \phi_{n \times p}(\mathbf{X}_i \mid \mathbf{M}_g + W_{ig} \mathbf{A}_g + \boldsymbol{\Lambda}_g \boldsymbol{\mathcal{Y}}_{ig}^B, W_{ig} \boldsymbol{\Sigma}_g, \boldsymbol{\Psi}_g^*) 
\right] \\
&= \mathcal{C} + \sum_{i=1}^N \sum_{g=1}^G -\frac{1}{2} Z_{ig} \Bigg[
p \log |\boldsymbol{\Sigma}_g|
+ \operatorname{tr} \Bigg\{
\frac{1}{W_{ig}} \boldsymbol{\Sigma}_g^{-1} (\mathbf{X}_i - \mathbf{M}_g)(\boldsymbol{\Psi}_g^*)^{-1}(\mathbf{X}_i - \mathbf{M}_g)'  \\
&\quad- 2\boldsymbol{\Sigma}_g^{-1} (\mathbf{X}_i - \mathbf{M}_g)(\boldsymbol{\Psi}_g^*)^{-1} \mathbf{A}_g'
- \frac{2}{W_{ig}} \boldsymbol{\Sigma}_g^{-1} (\mathbf{X}_i - \mathbf{M}_g)(\boldsymbol{\Psi}_g^*)^{-1} \boldsymbol{\mathcal{Y}}_{ig}^{B\prime} \boldsymbol{\Lambda}_g \\
&\quad+2\boldsymbol{\Sigma}_g^{-1} \mathbf{A}_g (\boldsymbol{\Psi}_g^*)^{-1} \boldsymbol{\mathcal{Y}}_{ig}^{B\prime} \boldsymbol{\Lambda}^{\prime}_g
+W_{ig} \boldsymbol{\Sigma}_g^{-1} \mathbf{A}_g (\boldsymbol{\Psi}_g^*)^{-1} \mathbf{A}_g'
 \\
&\quad+ \frac{1}{W_{ig}} \boldsymbol{\Sigma}_g^{-1} \boldsymbol{\Lambda}_g \boldsymbol{\mathcal{Y}}_{ig}^B (\boldsymbol{\Psi}_g^*)^{-1} \boldsymbol{\mathcal{Y}}_{ig}^{B\prime} \boldsymbol{\Lambda}_g'
\Bigg\} \Bigg].
\end{aligned}
$$
\cite{gallaughermcnicholas2019} discusses the conditional distribution of $\boldsymbol{\mathcal{Y}}_{ig}^B$ given $\mathbf{X}_i$, $W_{ig}$, and component membership for the purposes of establishing the E-step of this stage. For this step, the following conditional expectations are calculated as:
$$
\begin{aligned}
\mathbf{E}^{(2)}_{1ig} 
&:= \mathbb{E}\!\left[\boldsymbol{\mathcal{Y}}^{B}_{ig} \,\middle|\, \hat{\boldsymbol{\vartheta}}, \mathbf{X}_i, Z_{ig}=1\right] 
= \mathbf{L}_g \left( \mathbf{X}_i - \hat{\mathbf{M}}^{(t + 1)}_g - a_{ig}^{(t + 1)}\hat{\mathbf{A}}^{(t + 1)}_g \right),
\\
\mathbf{E}^{(2)}_{2ig} 
&:= \mathbb{E}\!\left[ \frac{1}{W_{ig}} \, \boldsymbol{\mathcal{Y}}^{B}_{ig} \,\middle|\, \hat{\boldsymbol{\vartheta}}, \mathbf{X}_i, Z_{ig}=1 \right] 
= \mathbf{L}_g \left( b_{ig}^{(t + 1)}\left(\mathbf{X}_i - \hat{\mathbf{M}}^{(t + 1)}_g\right) - \hat{\mathbf{A}}^{(t + 1)}_g \right),
\\
\mathbf{E}^{(2)}_{3ig} 
&:= \mathbb{E}\!\left[ \frac{1}{W_{ig}} \, \boldsymbol{\mathcal{Y}}^{B}_{ig} \left(\hat{\boldsymbol{\Psi}}_g^*\right)^{-1} \boldsymbol{\mathcal{Y}}^{B\prime}_{ig} \,\middle|\, \hat{\boldsymbol{\vartheta}}, \mathbf{X}_i, Z_{ig}=1 \right] = p \left( \mathbf{I}_q + \hat{\boldsymbol{\Lambda}}_g^{\prime (t)} \left(\hat{\boldsymbol{\Sigma}}_g^{(t)}\right)^{-1} \hat{\boldsymbol{\Lambda}}_g^{(t)} \right)^{-1} 
\\[1.2ex]
&
+ b_{ig}^{(t + 1)} \, \mathbf{L}_g \left(\mathbf{X}_i - \hat{\mathbf{M}}^{(t+1)}_g\right)\left( \hat{\boldsymbol{\Psi}}_g^*\right)^{-1} \left(\mathbf{X}_i - \hat{\mathbf{M}}^{(t+1)}_g\right)^{\prime} \mathbf{L}_g^{\prime}, \\[1.2ex]
&- \mathbf{L}_g \left( \left(\mathbf{X}_i - \hat{\mathbf{M}}^{(t + 1)}_g\right)\left(\hat{\boldsymbol{\Psi}}_g^*\right)^{-1}\hat{\mathbf{A}}^{\prime(t + 1)}_g
+ \hat{\mathbf{A}}^{(t + 1)}_g \left(\hat{\boldsymbol{\Psi}}_g^*\right)^{-1} \left(\mathbf{X}_i - \hat{\mathbf{M}}^{(t + 1)}_g\right)^{\prime} \right) \mathbf{L}_g^{\prime} \\
&+ a_{ig}^{(t + 1)} \, \mathbf{L}_g \hat{\mathbf{A}}^{(t + 1)}_g \left(\hat{\boldsymbol{\Psi}}_g^*\right)^{-1} \hat{\mathbf{A}}_g^{\prime (t + 1)} \mathbf{L}_g^{\prime}
\end{aligned}
$$
where
$$
\begin{aligned}
\mathbf{L}_g = \left( \mathbf{I}_q + \hat{\boldsymbol{\Lambda}}^{\prime(t)}_g \left(\hat{ \boldsymbol{\Sigma}}_g^{(t)}\right)^{-1} \hat{\boldsymbol{\Lambda}}^{(t)}_g \right)^{-1} \hat{\boldsymbol{\Lambda}}_g^{\prime (t)} \left(\hat{ \boldsymbol{\Sigma}}_g^{(t)}\right)^{-1}.
\end{aligned}
$$

\subsection*{Stage 2 updates}\label{row-ccc-update}

For a constrained model with unconstrained column factor loading matrices ($\boldsymbol{\Lambda}_g$), the update will be the same regardless of the constraints places on $\boldsymbol{\Sigma}_g$.
{
\begin{equation*}
\hat{\boldsymbol{\Lambda}}_g^{(t+1)} 
\!= \!\left( \sum_{i=1}^{N} \hat{z}^{(t+1)}_{ig} \!
\left[ \left(\mathbf{X}_i - \hat{\mathbf{M}}^{(t+1)}_g\right)\!\left(\hat{ \!\boldsymbol{\Psi}}_{\!g}^*\!\right)^{\!\!-1}\!\!\mathbf{E}^{(2)\prime}_{2ig} 
- \hat{\mathbf{A}}^{(t+1)}_g \left(\hat{ \!\boldsymbol{\Psi}}_{\!g}^*\!\right)^{\!\!-1}\!\!\mathbf{E}^{(2)\prime}_{1ig} \right] \right)
\left( \sum_{i=1}^{N} \hat{z}^{(t+1)}_{ig}\mathbf{E}^{(2)}_{3ig} \right)^{-1}.
\end{equation*}
}
However, if $\boldsymbol{\Lambda}_g$ is constrained across components, then it's updating formula depends on the structure of $\boldsymbol{\Sigma}_g$. For constrained models whose row model are of the form ``CCC" or ``CCU" ($\boldsymbol{\Lambda}_g = \boldsymbol{\Lambda}$ and either $\boldsymbol{\Sigma}_g = \boldsymbol{\Sigma}$ or $\boldsymbol{\Sigma}_g = \sigma \boldsymbol{I}_n$), the updates for the constrained column factor loadings are:
{
\begin{equation*}
\hat{ \boldsymbol{\Lambda}}^{(t+1)} \!= \! \left(\sum_{i = 1}^N \sum_{g = 1}^G \hat{z}^{(t+1)}_{ig}\! \left[\!\left( \mathbf{X}_i - \hat{\mathbf{M}}^{(t+1)}_g\right) \!\left( \!\hat{\boldsymbol{\Psi}}^*_{\!g}\!\right)^{\!\!-1} \!\!\mathbf{E}^{(2)\prime}_{2ig} - \hat{\mathbf{A}}^{(t+1)}_g \left(\!\hat{\boldsymbol{\Psi}}^*_{\!g}\!\right)^{\!\!-1}\!\!\mathbf{E}^{(2)\prime}_{1ig} \right]\!
\right) \!\!\left( \sum_{i = 1}^N \sum_{g = 1}^G \hat{z}^{(t+1)}_{ig} \mathbf{E}^{(2)}_{3ig}\!\right)^{\!\!\!-1}\!\!\!.
\end{equation*}
}
For constrained models with a row model of the form ``CUC" (i.e. $\boldsymbol{\Lambda}_g = \boldsymbol{\Lambda}$ and $\boldsymbol{\Sigma}_g = \sigma_g \boldsymbol{I}_n$), the updates for these constrained models are:
{
\small
\begin{equation*}
\hat{ \boldsymbol{\Lambda}}^{(t + 1)} \!=  \!\left( \sum_{g = 1}^G \frac{1}{\hat{\sigma}^{(t)}_g}\!\sum_{i = 1}^N \hat{z}^{(t+1)}_{ig}\! \left[\left( \mathbf{X}_i - \hat{\mathbf{M}}^{(t+1)}_g\right) \!\left( \hat{\boldsymbol{\Psi}}^*_g\right)^{-1} \mathbf{E}^{(2)\prime}_{2ig} - \hat{\mathbf{A}}^{(t+1)}_g \left(\hat{ \boldsymbol{\Psi}}_g^*\right)^{-1}\mathbf{E}^{(2)\prime}_{1ig} \right]\!
\right) \!\!\left(  \sum_{g = 1}^G \frac{1}{\hat{\sigma}^{(t)}_g} \!\sum_{i = 1}^N\hat{z}^{(t+1)}_{ig} \mathbf{E}^{(2)}_{3ig}\!\right)^{\!\!\!-1}\!\!\!\!\!.
\end{equation*}
}
For constrained models with a row model of the form ``CUU" (i.e. $\boldsymbol{\Lambda}_g = \boldsymbol{\Lambda}$ and $\boldsymbol{\Sigma}_g = \boldsymbol{\Sigma}_g$), the updates for the row factor loading matrix must be
calculated by row. The $j$\textsuperscript{th} row of $\boldsymbol{\Lambda}$ can be
found as:
{
\small
\begin{equation*}
\hat{ \boldsymbol{\Lambda}}_{(j)}^{(t + 1)} \!= \!\! \left( \sum_{g = 1}^G \frac{1}{\hat{\sigma}^{(t)}_{g_{(jj)}}}\!\sum_{i = 1}^N \hat{z}^{(t + 1)}_{ig}\!\! \left[\!\left( \mathbf{X}_i - \hat{\mathbf{M}}^{(t + 1)}_g\right)\! \left( \!\hat{\boldsymbol{\Psi}}^*_g\!\right)^{\!\!-1} \!\!\mathbf{E}^{(2)\prime}_{2ig} - \hat{\mathbf{A}}^{(t + 1)}_g \left(\!\hat{ \boldsymbol{\Psi}}_g^*\!\right)^{-1}\!\!\mathbf{E}^{(2)\prime}_{1ig} \right]
\right)_{\!\!\!(j)}\!\! \left(  \sum_{g = 1}^G \frac{1}{\hat{\sigma}^{(t)}_{g_{(jj)}}} \!\sum_{i = 1}^N\hat{z}^{(t + 1)}_{ig} \mathbf{E}^{(2)}_{3ig}\!\right)^{\!\!-1}\!\!\!\!\!,
\end{equation*}
}
for $j = 1, \ldots, n$, where $\hat{\sigma}^{(t)}_{g_{(jj)}}$ is the $j$\textsuperscript{th} diagonal entry of $\hat{\boldsymbol{\Sigma}}^{(t)}_g$.

After updating the column factor loadings matrix, $\boldsymbol{\Sigma}_g$ is updated next in this stage. There are four possible updates for $\boldsymbol{\Sigma}_g$ which all utilize the diagonal entries of the matrices:
$$
\begin{aligned}
\mathbf{S}_g^L &:= \frac{1}{p} \sum_{i = 1}^N \hat{z}^{(t+ 1)}_{ig} \!\bigg[ b^{(t+ 1)}_{ig} \!\left(\mathbf{X}_i \!-\! \hat{\mathbf{M}}^{(t+ 1)}_g\right)\left(\hat{\boldsymbol{\Psi}}_g^*\right)^{-1}\!\left(\mathbf{X}_i \!-\! \hat{\mathbf{M}}^{(t+ 1)}_g\right)' \\
&\quad-\left(\hat{\mathbf{A}}^{(t+ 1)}_g + \hat{\boldsymbol{\Lambda}}^{(t+ 1)}_g \mathbf{E}^{(2)}_{2ig}\right)\left(\!\hat{\boldsymbol{\Psi}}_{\!g}^*\right)^{\!\!-1} \!\left(\mathbf{X}_i \!-\! \hat{\mathbf{M}}^{(t + 1)}_g\right)' 
+\hat{\boldsymbol{\Lambda}}^{(t+ 1)}_g \mathbf{E}^{(2)}_{1ig} \left(\!\hat{\boldsymbol{\Psi}}_{\!g}^*\right)^{\!\!-1} \hat{\mathbf{A}}^{\prime (t + 1)}_g 
\\&\quad-\! \left(\mathbf{X}_i \!-\! \hat{\mathbf{M}}^{(t + 1)}_g\right)\left(\!\hat{\boldsymbol{\Psi}}_{\!g}^*\right)^{\!\!-1} \left(\hat{\mathbf{A}}^{\prime(t + 1)}_g + \mathbf{E}^{(2)\prime}_{2ig} \hat{\boldsymbol{\Lambda}}_g^{\prime (t + 1)} \right) +\!\hat{\mathbf{A}}^{(t + 1)}_g \!\left(\!\hat{\boldsymbol{\Psi}}_{\!g}^*\right)^{\!\!-1} \!\!\mathbf{E}^{(2)\prime}_{1ig} \hat{\boldsymbol{\Lambda}}_g^{\prime (t + 1)}
\\&\quad+ a_{ig}^{(t + 1)}  \hat{\mathbf{A}}^{(t + 1)}_g \left(\!\hat{\boldsymbol{\Psi}}_{\!g}^*\right)^{\!\!-1} \hat{\mathbf{A}}^{\prime(t + 1)}_g +   \hat{\boldsymbol{\Lambda}}^{(t + 1)}_g \mathbf{E}^{(2)}_{3ig} \hat{\boldsymbol{\Lambda}}_g^{\prime (t + 1)} \bigg].
\end{aligned}
$$
There are eight possible row models for $\boldsymbol{\Sigma}_g^*$. The update for $\boldsymbol{\Sigma}_g$ is independent of the constraint on $\boldsymbol{\Lambda}_g$, so the updates are categorized in the following table:

\begin{table}[h!]
\centering
\caption{Updates to $\boldsymbol{\Sigma}_g$ for each row model}
\begin{tabular}{llll}
\toprule
Constraint & Isotropy & Row Model & Update \\
\midrule

\multirow{2}{*}{$\boldsymbol{\Sigma}_g$ unconstrained}
& Isotropic 
& CUC, UUC 
& $\hat{\sigma}_g^{(t+1)} = \frac{1}{N_g n} \mathrm{tr}(\mathbf{S}_g^L)$ \\[1.2ex]

& Anisotropic 
& CUU, UUU 
& $\hat{\boldsymbol{\Sigma}}_g^{(t+1)} = \frac{1}{N_g} \mathrm{diag}(\mathbf{S}_g^L)$ \\

\midrule

\multirow{2}{*}{$\boldsymbol{\Sigma}_g = \boldsymbol{\Sigma}$}
& Isotropic 
& CCC, UCC 
& $\hat{\sigma}^{(t+1)} = \frac{1}{N n} \sum_{g=1}^G \mathrm{tr}(\mathbf{S}_g^L)$ \\[1.2ex]

& Anisotropic 
& CCU, UCU 
& $\hat{\boldsymbol{\Sigma}}^{(t+1)} = \frac{1}{N} \sum_{g=1}^G \mathrm{diag}(\mathbf{S}_g^L)$ \\

\bottomrule
\end{tabular}
\end{table}

\subsection*{AECM stage 3}

For the calculation of $\boldsymbol{\Psi}_g$ and $\boldsymbol{\Delta}_g$ for $g = 1,\ldots, G$, we consider the complete log-likelihood with the same information as in stage 1 with the addition of $\boldsymbol{\mathcal{Y}}_{ig}^A$ for $g = 1,\ldots, G$ for each $i = 1\,\ldots,N$. This complete log-likelihood can be expressed as:
$$
\begin{aligned}
\ell_{\mathrm{C}_3} &=\; C + \sum_{i=1}^N \sum_{g=1}^G Z_{ig} \Bigg[ \log \pi_g + \log h(W_{ig} \mid \boldsymbol{\theta}_g) + \log \phi_{q \times p}(\mathcal{Y}_{ig}^A \mid \boldsymbol{0}, W_{ig} \boldsymbol{\Sigma}_g^*, \mathbf{I}_p) \\
&\quad + \log \phi_{n \times p}(\mathbf{X}_i \mid \mathbf{M}_g + W_{ig} \mathbf{A}_g + \mathcal{Y}_{ig}^A \boldsymbol{\Delta}_g^\prime, W_{ig} \boldsymbol{\Sigma}_g^*, \boldsymbol{\Psi}_g) \Bigg] \\
&=\; C + \sum_{i=1}^N \sum_{g=1}^G -\frac{1}{2} Z_{ig} \Bigg[ n \log |\boldsymbol{\Psi}_g| + \operatorname{tr} \Bigg\{ \frac{1}{W_{ig}} \boldsymbol{\Psi}_g^{-1} (\mathbf{X}_i - \mathbf{M}_g)' (\boldsymbol{\Sigma}_g^*)^{-1} (\mathbf{X}_i - \mathbf{M}_g) \\
&\quad - 2\boldsymbol{\Psi}_g^{-1} (\mathbf{X}_i - \mathbf{M}_g)' (\boldsymbol{\Sigma}_g^*)^{-1} \mathbf{A}_g - \frac{2}{W_{ig}} \boldsymbol{\Psi}_g^{-1} (\mathbf{X}_i - \mathbf{M}_g)' (\boldsymbol{\Sigma}_g^*)^{-1} \mathcal{Y}_{ig}^A \boldsymbol{\Delta}_g^\prime \\
&\quad + 2\boldsymbol{\Psi}_g^{-1} \mathbf{A}_g^\prime (\boldsymbol{\Sigma}_g^*)^{-1} \mathcal{Y}_{ig}^A \boldsymbol{\Delta}_g^\prime + W_{ig} \boldsymbol{\Psi}_g^{-1} \mathbf{A}_g^\prime (\boldsymbol{\Sigma}_g^*)^{-1} \mathbf{A}_g  \\
&\quad + \frac{1}{W_{ig}} \boldsymbol{\Psi}_g^{-1} \boldsymbol{\Delta}_g \mathcal{Y}_{ig}^{A\prime} (\boldsymbol{\Sigma}_g^*)^{-1} \mathcal{Y}_{ig}^A \boldsymbol{\Delta}_g^\prime \Bigg\} \Bigg].
\end{aligned}
$$
Before performing the calculations in this stage, it should be noted that the scale matrix $\hat{\boldsymbol{\Sigma}}_g^*$ should be updated with the newly updated values, $\hat{\boldsymbol{\Lambda}}^{(t + 1)}_g$ and $\hat{\boldsymbol{\Sigma}}^{(t+1)}_g$, from stage 2. Like in the previous stage,
\cite{gallaughermcnicholas2019} also mentions the conditional distribution of $\boldsymbol{\mathcal{Y}}_{ig}^A$ given $\mathbf{X}_i$, $W_{ig}$, component membership, and parameter updates from stage 1 and 2 for the purposes of establishing the E-step of this stage. For this step, the following conditional expectations are calculated as:
$$
\begin{aligned}
\mathbf{E}^{(3)}_{1ig} 
&:= \mathbb{E}\!\left[\mathcal{Y}^{A}_{ig} \,\middle|\, \hat{\boldsymbol{\vartheta}}, \mathbf{X}_i, Z_{ig}=1\right] 
= \left(\mathbf{X}_i - \hat{\mathbf{M}}^{(t + 1)}_g - a_{ig}^{(t + 1)}\hat{\mathbf{A}}^{(t + 1)}_g\right)\mathbf{D}_g,
\\
\mathbf{E}^{(3)}_{2ig} 
&:= \mathbb{E}\!\left[ \frac{1}{W_{ig}} \, \mathcal{Y}^{A}_{ig} \,\middle|\, \hat{\boldsymbol{\vartheta}}, \mathbf{X}_i, Z_{ig}=1 \right] 
= \left( b_{ig}^{(t + 1)}\left(\mathbf{X}_i - \hat{\mathbf{M}}^{(t + 1)}_g\right) - \hat{\mathbf{A}}^{(t + 1)}_g \right)\mathbf{D}_g, 
\\
\mathbf{E}^{(3)}_{3ig} 
&:= \mathbb{E}\!\left[ \frac{1}{W_{ig}} \, \mathcal{Y}^{A\prime}_{ig} \left(\!\hat{ \boldsymbol{\Sigma}}_g^*\!\right)^{\!\!-1} \mathcal{Y}^{A}_{ig} \,\middle|\, \hat{\boldsymbol{\vartheta}}, \mathbf{X}_i, Z_{ig}=1 \right] = n\left(\mathbf{I}_r + \hat{ \boldsymbol{\Delta}}_g^{\!\prime (t)}\left(\!\hat{ \boldsymbol{\Psi}}_{g}^{(t)}\!\right)^{\!\!-1}\!\hat{ \boldsymbol{\Delta}}^{(t)}_g \right)^{-1} 
\\&\quad+ b_{ig}^{(t + 1)}\mathbf{D}_g^{\prime}\left(\mathbf{X}_i - \hat{\mathbf{M}}^{(t + 1)}_g\right)^{\prime}\left(\!\hat{ \boldsymbol{\Sigma}}_g^*\!\right)^{\!\!-1}\left(\mathbf{X}_i - \hat{\mathbf{M}}^{(t + 1)}_g\right)\mathbf{D}_g \\
&\quad - \mathbf{D}_g^{\prime} \bigg(\!\left(\mathbf{X}_i - \hat{\mathbf{M}}^{(t + 1)}_g\right)^{\prime}\left(\!\hat{ \boldsymbol{\Sigma}}_g^*\!\right)^{\!-1}\!\hat{\mathbf{A}}^{(t+1)}_g + \hat{\mathbf{A}}_g^{\!\prime (t + 1)}\left(\!\hat{ \boldsymbol{\Sigma}}_g^*\!\right)^{\!\!-1}\!\left(\mathbf{X}_i - \hat{\mathbf{M}}^{(t + 1)}_g\right)\!\bigg)\mathbf{D}_g \\
&\quad + a^{(t + 1)}_{ig}\mathbf{D}_g^{\prime}\hat{\mathbf{A}}_g^{\!\prime (t + 1)}\left(\!\hat{ \boldsymbol{\Sigma}}_g^*\!\right)^{\!-1}\!\hat{\mathbf{A}}^{(t+1)}_g\mathbf{D}_g
\end{aligned}
$$
where
$$
\begin{aligned}
\mathbf{D}_g = \left(\!\hat{ \boldsymbol{\Psi}}_{g}^{(t)}\!\right)^{\!\!-1}\!\hat{ \boldsymbol{\Delta}}^{(t)}_g\left(\mathbf{I}_r + \hat{ \boldsymbol{\Delta}}_g^{\!\prime (t)}\left(\!\hat{ \boldsymbol{\Psi}}_{g}^{(t)}\!\right)^{\!\!-1}\!\hat{ \boldsymbol{\Delta}}_g^{(t)}\right)^{-1}.
\end{aligned}
$$

\subsection*{Stage 3 updates}

For a constrained model with unconstrained row factor loading matrices ($\boldsymbol{\Delta}_g$), the update will be the same regardless of the constraints places on $\boldsymbol{\Psi}_g$. The row factor loading matrices are updated as:
$$
\begin{aligned}
\hat{\boldsymbol{\Delta}}^{(t + 1)}_g 
= \left( \sum_{i=1}^{N} \hat{z}^{(t + 1)}_{ig}\! 
\Big[ \!\left(\mathbf{X}_i - \hat{\mathbf{M}}^{(t + 1)}_g\right)^{\prime}\left(\!\boldsymbol{\hat{\Sigma}}^{*}_g\!\right)^{\!-1}\!\mathbf{E}^{(3)}_{2ig} 
- \hat{\mathbf{A}}_g^{\!\prime (t + 1)}\left(\!\boldsymbol{\hat{\Sigma}}^{*}_g\!\right)^{\!-1}\!\mathbf{E}^{(3)}_{1ig} \Big] \right)\!\!
\left( \sum_{i=1}^{N} \hat{z}^{(t + 1)}_{ig}\mathbf{E}^{(3)}_{3ig} \right)^{-1}
\end{aligned}
$$

However, similar to the stage 2 updates, if $\boldsymbol{\Delta}_g$ is constrained across components, then it's updating formula depends on the structure of $\boldsymbol{\Psi}_g$. For constrained models whose column model are of the form ``CCC" or ``CCU" ($\boldsymbol{\Delta}_g = \boldsymbol{\Delta}$ and either $\boldsymbol{\Psi}_g = \boldsymbol{\Psi}$ or $\boldsymbol{\Psi}_g = \psi \boldsymbol{I}_p$), the update for the constrained row factor loading matrix is:
$$
\begin{aligned}
\hat{\boldsymbol{\Delta}} &= \left( \sum_{i = 1}^N \sum_{g = 1}^G \hat{z}^{(t + 1)}_{ig}\!\! \left[ \left( \mathbf{X}_i - \hat{\mathbf{M}}^{(t + 1)}_g \right)^{\prime} \left(\!\boldsymbol{\hat{\Sigma}}^{*}_g\!\right)^{\!-1}\! \mathbf{E}^{(3)}_{2ig} - \hat{\mathbf{A}}_{g}^{\!\prime(t+1)} \left(\!\boldsymbol{\hat{\Sigma}}^{*}_g\!\right)^{\!-1}\! \mathbf{E}^{(3)}_{1ig}\right]\right)\!\! \left( \sum_{i = 1}^N \sum_{g = 1}^G \hat{z}^{(t + 1)}_{ig} \mathbf{E}^{(3)}_{1ig} 
\right)^{\!\!-1}\!\!\!.
\end{aligned}
$$

For constrained models with a column model of the form ``CUC" (i.e. $\boldsymbol{\Delta}_g = \boldsymbol{\Delta}$ and $\boldsymbol{\Psi}_g = \psi_g \boldsymbol{I}_p$), the row factor loading update for this column model is:
$$
\begin{aligned}
\hat{\boldsymbol{\Delta}} &= \left( \sum_{g = 1}^G \frac{1}{\hat{\psi}^{(t)}_g}\!\!\sum_{i = 1}^N \hat{z}^{(t + 1)}_{ig}\!\! \left[ \left( \mathbf{X}_i \!- \!\hat{\mathbf{M}}^{(t + 1)}_g \right)^{\prime} \!\left(\!\boldsymbol{\hat{\Sigma}}^{*}_g\!\right)^{\!-1}\!\! \mathbf{E}^{(3)}_{2ig} \!-\! \hat{\mathbf{A}}_{g}^{\prime} \left(\!\boldsymbol{\hat{\Sigma}}^{*}_g\!\right)^{\!-1}\!\! \mathbf{E}^{(3)}_{1ig}\right]\right) \!\!\left( \sum_{g = 1}^G \frac{1}{\hat{\psi}^{(t)}_g} \!\!\sum_{i = 1}^N  \hat{z}^{(t + 1)}_{ig} \mathbf{E}^{(3)}_{1ig} 
\!\right)^{\!\!-1}\!\!\!,
\end{aligned}
$$

For constrained models with a column model of the form ``CUU" (i.e. $\boldsymbol{\Delta}_g = \boldsymbol{\Delta}$ and $\boldsymbol{\Psi}_g = \boldsymbol{\Psi}_g$), the updates for the row factor loading matrix must be
calculated row by row. The $j$th row of $\boldsymbol{\Delta}$ can be
found as:
{
\small
\begin{equation*}
\hat{\boldsymbol{\Delta}}_{(j)} \!= \!\!\left( \sum_{g = 1}^G \frac{1}{\hat{\psi}^{(t)}_{g_{(jj)}}}\!\!\sum_{i = 1}^N  \hat{Z}_{ig}^{(t + 1)}\!\! \left[ \left( \mathbf{X}_i \!-\! \hat{\mathbf{M}}^{(t + 1)}_g \right)^{\prime}\! \left(\!\boldsymbol{\hat{\Sigma}}^{*}_g\!\right)^{\!-1}\!\! \mathbf{E}^{(3)}_{2ig} \!-\! \hat{\mathbf{A}}_{g}^{\!\prime (t + 1)}  \!\left(\!\boldsymbol{\hat{\Sigma}}^{*}_g\!\right)^{\!-1}\!\!\mathbf{E}^{(3)}_{1ig}\right]\!\right)_{\!\!\!(j)} \!\!\left( \sum_{g = 1}^G \frac{1}{\hat{\psi}^{(t)}_{g_{(jj)}}} \!\!\sum_{i = 1}^N \hat{z}^{(t + 1)}_{ig} \mathbf{E}^{(3)}_{1ig} 
\right)^{\!\!\!-1}\!\!\!\!,
\end{equation*}
}

for $j = 1, \ldots, p$, where $\hat{\psi}^{(t)}_{g_{(jj)}}$ is the $j$th diagonal entry of $\hat{\boldsymbol{\Psi}}_g$.

After updating the row factor loadings matrix, $\boldsymbol{\Psi}_g$ is updated next in this stage. There are four possible updates for $\boldsymbol{\Psi}_g$ which all utilize the diagonal entries of the matrices:
\begin{equation*}
\begin{aligned}
\mathbf{S}^D_g 
&:= \frac{1}{ n} \sum_{i=1}^N \hat{z}^{(t + 1)}_{ig} \Bigg[
b_{ig}^{(t + 1)}\left(\mathbf{X}_i - \hat{\mathbf{M}}^{(t + 1)}_g\right)^{\prime}\!\left(\boldsymbol{\hat{\Sigma}}^{*}_g\right)^{\!-1}\!
\!\left(\mathbf{X}_i - \hat{\mathbf{M}}^{(t + 1)}_g\right) \\
&\quad-\left(\hat{\mathbf{A}}_g^{\!\prime (t + 1)} + \hat{\boldsymbol{\Delta}}^{(t + 1)}_g \mathbf{E}^{(3)\prime}_{2ig}\right)\left(\boldsymbol{\hat{\Sigma}}^{*}_g\right)^{\!-1}\!\!\left(\mathbf{X}_i - \hat{\mathbf{M}}^{(t + 1)}_g\right) + \hat{\boldsymbol{\Delta}}^{(t + 1)}_g \mathbf{E}^{(3)\prime}_{1ig}\left(\boldsymbol{\hat{\Sigma}}^{*}_g\right)^{-1}\!\hat{\mathbf{A}}^{\!(t + 1)}_g\\
&\quad- \left(\mathbf{X}_i - \hat{\mathbf{M}}^{(t + 1)}_g\right)^{\prime}\left(\boldsymbol{\hat{\Sigma}}^{*}_g\right)^{-1}\left(\hat{\mathbf{A}}^{\!(t + 1)}_g + \mathbf{E}^{(3)}_{2ig}\hat{\boldsymbol{\Delta}}_g^{\!\prime (t + 1)} \right)
+\hat{\mathbf{A}}_g^{\!\prime(t + 1)} \left(\hat{ \boldsymbol{\Sigma}}_g^*\right)^{\!\!-1}\! \mathbf{E}^{(3)}_{1ig} \hat{\boldsymbol{\Delta}}_g^{\!\prime(t + 1)}
\\
&\quad
+ a_{ig}^{(t +1)}\hat{\mathbf{A}}_g^{\!\prime (t + 1)}\left(\boldsymbol{\hat{\Sigma}}^{*}_g\right)^{-1}\hat{\mathbf{A}}^{\!(t + 1)}_g 
+ \hat{\boldsymbol{\Delta}}_g^{\!(t + 1)} \mathbf{E}^{(3)}_{3ig}\hat{\boldsymbol{\Delta}}_g^{\!\prime(t + 1)}
\Bigg].
\end{aligned}
\end{equation*}
Like the row models, there are eight possible column models for $\boldsymbol{\Psi}_g^*$. The update for $\boldsymbol{\Psi}_g$ is independent of the constraint on $\boldsymbol{\Delta}_g$, so the updates are categorized in the following table:

\begin{table}[h!]
\centering
\caption{Updates to $\boldsymbol{\Psi}_g$ for each column model}
\begin{tabular}{llll}
\toprule
Constraint & Structure & Column Model & Update \\
\midrule

\multirow{2}{*}{$\boldsymbol{\Psi}_g$ unconstrained}
& Isotropic 
& CUC, UUC 
& $\hat{\psi}_g^{(t+1)} = \frac{1}{N_g p} \mathrm{tr}(\mathbf{S}_g^D)$ \\[1.2ex]

& Anisotropic 
& CUU, UUU 
& $\hat{\boldsymbol{\Psi}}_g^{(t+1)} = \frac{1}{N_g} \mathrm{diag}(\mathbf{S}_g^D)$ \\

\midrule

\multirow{2}{*}{$\boldsymbol{\Psi}_g = \boldsymbol{\Psi}$}
& Isotropic 
& CCC, UCC 
& $\hat{\psi}^{(t+1)} = \frac{1}{N p} \sum_{g=1}^G \mathrm{tr}(\mathbf{S}_g^D)$ \\[1.2ex]

& Anisotropic 
& CCU, UCU 
& $\hat{\boldsymbol{\Psi}}^{(t+1)} = \frac{1}{N} \sum_{g=1}^G \mathrm{diag}(\mathbf{S}_g^D)$ \\

\bottomrule
\end{tabular}
\end{table}

\subsection{Initialization and convergence criterion}

\subsection*{Initialization details}

The following section lists out recommendations for model initializations that experiments on simulated and real data suggest are stable and converge to an optimal model. For clustering, Every model fit will try many ``candidate" initializations of $\hat{\boldsymbol{\vartheta}}^{(0)}$. The candidates are generated as follows:

The algorithm starts off by assuming a balanced sample among $G$ clusters (i.e. $\hat{\pi}^{(0)}_g = 1 / G$). The location matrix for the $g$th group, $\mathbf{M}_g$, is initialized similarly to the initialization of multivariate $k$-means clustering. Starting with $g = 1$, select an observation at random, otherwise, select an observation which hasn't been placed into one of the previously created groups. This observation will be referred to as $\tilde{ \mathbf{X}}_g$. Next, calculate $\|\tilde{ \mathbf{X}}_g - \mathbf{X}_i\|_F$ for $i = 1,\ldots,N$, and order the observations by their distance from $\tilde{\mathbf{X}}_g$ in the Frobenius norm in ascending order. Lastly, select the first approximately $N / G$ observations with the smallest Frobenius norm quantity out of the observations that have not already been assigned to a previous group. Only the selected observations will contribute to the calculation of the $g$\textsuperscript{th} location matrix. In other words, $\hat{ \mathbf{M}}_g^{(0)} = \frac{1}{N_g}\sum_{i = 1}^N \hat{z}^{(0)}_{ig} \mathbf{X}_i$, where $N_g = \sum_{i = 1}^N \hat{z}^{(0)}_{ig}$ and $\hat{z}^{(0)}_{ig} = 1$ if observation $i$ was assigned to group $g$ and $\hat{z}^{(0)}_{ig} = 0$ otherwise. On the other hand, the initializations of $\mathbf{A}_g$ don't appear to have much of an impact on stability or model fit. So, the skewness parameters are initialized to be: $\hat{\mathbf{A}}_g^{(0)} = 0.1 \times \boldsymbol{1}_n \boldsymbol{1}_p^{\prime}$.

The initializations for the column and row factor loading matrices, each element of $\hat{\boldsymbol{\Lambda}}^{(0)}_g$ and $\hat{\boldsymbol{\Delta}}^{(0)}_g$ are initialized with random draws from a  uniform distribution on $(-1, 1)$. The factor loading matrices for each component will also be given the same random initializations.

Lastly, the initializations of the diagonal matrix. Hence, $\hat{\boldsymbol{\Sigma}}_g^{(0)} = \boldsymbol{I}_n$ and $\hat{\boldsymbol{\Psi}}_g^{(0)} = \boldsymbol{I}_p$ for $g = 1,\ldots,G$. Additionally, the initialization of the concentration parameters are $\hat{\nu}^{(0)}_g = 10$.

\subsection*{Small EM - Big EM}

For models fits on data demonstrated in this paper are all fit with the following procedure.
First, two-hundred candidate initializations are formed, and ordered by their observed likelihood using \eqref{eq:oll}. Next, twenty of the candidates with the greatest observed log-likelihood are kept and run in the "small EM" part of the algorithm. 

These twenty candidate sets are then run through the AECM algorithm with the constrained model (CCC-CCC-U-C) for exactly four iterations, regardless of the actual constrained model being fit. Out of the twenty parameter sets, the set with the largest observed log-likelihood after the four iterations will be used to initialize the Big EM algorithm. The ``Big EM" here is the AECM algorithm iterated with updates defined by the desired constrained model, run to full convergence.

Empirical evidence suggests this approach to initialization yields stable results for every possible constrained model. Some simulations suggest leaving the skewness matrices unconstrained in the small EM step will improve classification in data sets with extreme outliers. However, if the Big EM step is iterated over a model with constrained skewness, using initializations with unconstrained skewness may sometimes result in the first iteration resulting in a decrease in the observed log-likelihood. After this one potential decrease, the parameter set, now having readjusted into the appropriate parameter space with constrained skewness, acts as a new initialization for the desired model and experiences a monotonically increasing $\ell \left(\hat{\boldsymbol{\vartheta}}\right)$ as expected.

\subsection*{Convergence criterion}

The models studied in this dataset rely on the relatively simple ``lack of progress" convergence criterion. For this paper, the AECM algorithm is said to have converged after $t+1$ iterations if 
\begin{equation}
\label{eq:conv-crit}
\ell \left(\hat{\boldsymbol{\vartheta}}^{(t + 1)}\right) - \ell \left(\hat{\boldsymbol{\vartheta}}^{(t)}\right) < \epsilon,
\end{equation}
for $\epsilon > 0$. Empirically, $\epsilon = 0.01$ is more than enough lack of progress when comparing likelihoods on a logarithm scale. However, reaching this tolerance level may take a burdensome amount of iterations before reaching convergence. So, this paper uses an $\epsilon$ that is dependent on certain features of the data and the model fit. The tolerance for model fits in this paper are determined as follows. Let $\ell^{(t)}$ be shorthand for $\ell \left(\hat{\boldsymbol{\vartheta}}^{(t)}\right)$. Then, the observed log-likelihood at the tenth iteration is denoted with $\ell^{(10)}$.  Then, $\epsilon = |\ell^{(10)}| \times 10^{-k}$ where $k = \log_{10} \left( np \sqrt{N}\right) + 3$. In simulation and real data experiments, this results in $\epsilon$ values that are on the order of $10^{-1}$ or $10^{-2}$. Using $\ell^{(10)}$ allows for slightly larger tolerances models that don't appear as likely after 10 iterations. This choice of $k$ hovers between 6 and 7.5 which usually makes the value of $10^{k}$ an order of magnitude or more larger than $\ell^{(10)}$. Unfortunately, at this level of tolerance, it may take thousands of iterations before successful convergence. However, experiments suggest this is an appropriate level of tolerance for accurate model recovery.

\subsection{Notes on identifiability}

Estimates of $\boldsymbol{\Sigma}^*_g$ and $\boldsymbol{\Psi}^*_g$ are only unique up to a positive constant. For a mixture model that has an unconstrained column model (i.e. not CCC or CCU) or if both row and column models are constrained, the identifiability issue can be resolved by imposing a restriction on $\boldsymbol{\Sigma}^*_g$. For example, the first diagonal entry of $\boldsymbol{\Sigma}^*_g$ can be restricted to be 1 for these particular constrained models as suggested in \cite{gallaugher18a}. However, if the column model is constrained (CCC or CCU) and the row model is not, a constraint should be imposed on the scale matrix that's shared among components ($\boldsymbol{\Psi}^*$) instead. 

The factor loading parameters $\boldsymbol{\Lambda}_g$ and $\boldsymbol{\Delta}_g$ are both only identifiable up to a rigid rotation. Suppose $\boldsymbol{U}$ is a $q \times q$ unitary matrix. Then using $\boldsymbol{\Lambda}_g^* := \boldsymbol{\Lambda}_g \boldsymbol{U}$ in place of $\boldsymbol{\Lambda}_g$ would have no effect on the model. Likewise, let $\boldsymbol{U}$ is a $r \times r$ unitary matrix. Then using $\boldsymbol{\Delta}_g^* := \boldsymbol{\Delta}_g \boldsymbol{U}$ in place of $\boldsymbol{\Delta}_g$ would have no effect on the model. \cite{lawley71} discusses how this fact affects the number of parameters that a factor loading matrix contributes to a model which is reflected in Table \ref{tab:rowmodels} and Table \ref{tab:colmodels}.

Lastly, like any clustering problem, the model is invariant upon relabeling components (discussed more in \cite{stephens00}). The label switching identifiability issue can cause problems for mixture models utilizing the Bayesian paradigm as Stephens discusses. However, it is not a practical concern for the parameter estimation methodology discussed in this paper. Computationally, randomly assigning labels will suffice for the purposes of clustering. Theoretically, the labels can be restricted to be ordered by the proportion of each component (I.e. $\pi_1 > \pi_2 > \ldots > \pi_G$) in order to resolve the label switching issue.

\subsection{Semi-supervised classification}

Clustering is the case where none of the labels of the observations are known. However, the mixture model discussed here can easily be extended to include cases for semi-supervised classification, the case when some but not all of the labels to the observations are known. Suppose, without loss of generality, that the labels of the first $K$ observations, ($\boldsymbol{Z}_{1}, \ldots, \boldsymbol{Z}_K$) are known. Then, the observed log-likelihood of the data can be expressed as:
$$
\begin{aligned}
\ell \left( \hat{\boldsymbol{\vartheta}}^{(t)} \mid \mathbf{X}\right) := \sum_{i=1}^{K} \log \left( \sum_{g=1}^G Z_{ig}\left[ \hat{\pi}^{(t)}_g f\left(\mathbf{X}_i \mid \hat{\boldsymbol{\vartheta}}_{g}^{(t)}\right)\right]\right) +
\sum_{i=K + 1}^{N} \!\!\log \left( \sum_{h=1}^H  \hat{\pi}^{(t)}_h f\left(\mathbf{X}_i \mid \hat{\boldsymbol{\vartheta}}_{h}^{(t)}\right)\right)
\end{aligned}
$$

where $H \geq G$ generally speaking. This paper however only discusses the more common cases where $H = G$. More details about mixture modeling with semi-supervised classification are discussed in \cite{mclachlan00b} and \cite{mcnicholas16a}.

\subsection{Model assessment}

For a general clustering problem, the number of components ($G$), number of row factors ($r$), number of column factors ($q$), and the constrained model ($M$) are not known in advance. Since $G$, $r$, and $q$ are not parameters that can be assessed through means of a hypothesis test, there are two main approaches to model selection that are common in the literature of mixture modeling. The Bayesian Information Criterion (\cite{schwarz78}) is a linear combination of the maximized observed log-likelihood ($\ell (\hat{\boldsymbol{\vartheta}})$) and the number of free parameters ($k (M)$) associated with the constrained model ($M$), defined as follows:
\begin{equation}
\label{eq:bic-def}
BIC (M) = -2\ell (\hat{\boldsymbol{\vartheta}}) + k (M) \log (N).
\end{equation}
This notation will be useful for describing the adaptive algorithm approach, an alternative to fitting all possible constrained models for given levels of $G$, $r$, and $q$. This paper uses the convention that a model with a lower BIC is preferred over one with a higher BIC. This information criterion is a useful tool for assessing model fit because unnecessary parameters will be heavily weighted in order to select more parsimonious models. Another commonly used information criterion is the integrated completed likelihood (ICL) (\cite{biernacki00}), which is approximated as:
$$
ICL(M) \approx BIC (M) - 2 \sum_{i = 1}^N \sum_{g = 1}^G  \operatorname{MAP} (\hat{Z}_{ig})\log \hat{Z}_{ig},\text{ where } \operatorname{MAP} (\hat{Z}_{ig}) 
= \begin{cases}
    1 & \text{if } \underset{h=1,\ldots,G}{\operatorname{arg max}} \,\hat{Z}_{ig} = 1 \\
    0 & \text{otherwise}
\end{cases}
$$

The ICL is essentially the BIC but with an additional penalty for models with more component membership uncertainty.

To assess clustering/classification performance, the adjusted Rand index is a useful tool for assessing class agreement even when classification labels are different. This measure is useful for many model-based clustering applications which can identifiability issues resulting from label switching. When comparing predicted classification with the true classifications, the ARI is 1 when there is perfect classification. The ARI is close to 0 when the predicted classification is no better than randomly predicting classification (\cite{steinley04}).  The misclassification rate (MCR), the number of misclassified observations divided by $N$, will also be used to more directly report model accuracy. 

\section{Numerical considerations}

This section covers various details surrounding computational and theoretical concerns as well as implementation and stability improvements for the algorithms discussed in the previous section.

\subsection{Existence and uniqueness of a constrained skewness parameter}

This section covers details regarding models in the PMSBFA family that constrain the skewness matrix across components. The update for the constrained skewness matrix should provide updates for $\mathbf{A}$ that are guaranteed to exist and are unique at each iteration. If we can show that the large matrix on the left hand side of \eqref{eq:const-skew-upd} is symmetric and positive definite, this is more than enough prove that the linear system in invertible and a unique solution for $\mathbf{A}$ exists. Define the following $np \times np$ matrix:
\begin{equation}
\label{eq:large-mat}
\mathcal{B} :=\sum_{g = 1}^G \!N_g \!\left( \bar{a}_g \!- \!\frac{1}{\bar{b}_g} \right) \!\!\left[ \left( \boldsymbol{\Psi}_{\!g}^*\right)^{\!\!-1} \!\!\otimes\! \left(\boldsymbol{\Sigma}_{g}^*\right)^{\!\!-1}\right]
\end{equation}

Firstly, both scale covariance matrices, $\boldsymbol{\Sigma}_g^*$ and $\boldsymbol{\Psi}_g^*$, are symmetric and positive definite (SPD) by definition. Therefore, $(\boldsymbol{\Sigma}_g^*)^{-1}$ and $(\boldsymbol{\Psi}_g^*)^{-1}$ are also SPD. Then, via properties of the Kronecker product, we also must have that $(\boldsymbol{\Psi}_g^*)^{-1} \otimes (\boldsymbol{\Sigma}_g^*)^{-1}$ is SPD. Lastly, a weighted sum of SPD matrices is also SPD if and only if the weights are all positive. The ``weights" needed to construct $\mathcal{B}$ are $N_g \left( \bar{a}_g - 1/\bar{b}_g \right)$ for $g = 1, \ldots, G$. It is not hard to establish that $N_g > 0$ since $\hat{Z}_{ig} > 0$ by definition \eqref{eq:z-update}. Also, $a_{ig}$ and $b_{ig}$ are positive by their definition \eqref{eq:cond-exp-def} since the support of $W_{ig}$ is strictly positive. However, it isn't as clear that $\bar{a}_g - 1/\bar{b}_g > 0$. To prove this, first consider the function $h(x) = 1/x$. It is well established that $h(x)$ is strictly convex for $x > 0$, and so Jensen's inequality can be used here. Utilizing the definition of $a_{ig}$ and $b_{ig}$ from \eqref{eq:cond-exp-def} yields the following inequality:
\begin{equation*}
\begin{aligned}
h \bigg( \mathbb{E} \!\bigg[ \!W_{ig} \bigg|\mathbf{X}_i, Z_{ig} \!=\! 1, \hat{\boldsymbol{\vartheta}}_g^{(t)} \!\bigg] \bigg) &< \mathbb{E} \!\bigg[ h \left( W_{ig}\right) \bigg|\mathbf{X}_i, Z_{ig} \!=\! 1, \hat{\boldsymbol{\vartheta}}_g^{(t)} \!\bigg],  \text{ (via Jensen's Inequality.)} \\
\implies \,\, \frac{1}{a_{ig}} &< b_{ig}, \text{ (via definitions from \eqref{eq:cond-exp-def})} \\
 a_{ig} > \frac{1}{b_{ig}} \implies \hat{Z}_{ig} a_{ig}  &> \hat{Z}_{ig}\frac{1}{b_{ig}} \implies \sum_{i = 1}^N \hat{Z}_{ig} a_{ig} > \sum_{i = 1}^N \hat{Z}_{ig}\frac{1}{b_{ig}} \\
 \therefore \, \bar{a}_g &> \frac{1}{N_g}\sum_{i = 1}^N \hat{Z}_{ig}\frac{1}{b_{ig}}.
\end{aligned}
\end{equation*}
Jensen's inequality also gives us this result for convex combinations of functions like $h$:
\begin{equation*}
h \left( \frac{1}{N_g} \sum_{i = 1}^N \hat{Z}_{ig} b_{ig}\right) < 
\frac{1}{N_g} \sum_{i = 1}^N \hat{Z}_{ig} h(b_{ig}) \implies
\frac{1}{\bar{b}_g} < \frac{1}{N_g} \sum_{i = 1}^N \hat{Z}_{ig}\frac{1}{b_{ig}}.
\end{equation*}
So, combining the two inequalities yields:
\begin{equation*}
\bar{a}_g > \frac{1}{N_g}\sum_{i = 1}^N \hat{Z}_{ig}\frac{1}{b_{ig}} > \frac{1}{\bar{b}_g} \implies \bar{a}_g > \frac{1}{\bar{b}_g}.
\end{equation*}
Hence, we have that $\bar{a}_g - 1/\bar{b}_g$ is positive. Therefore, we can conclude that $\mathcal{B}$ is symmetric and positive definite. Thus, $\mathcal{B}$ is invertible, and a unique solution for $\hat{\mathbf{A}}$ must exist.

\subsection{The Conjugate Gradient method with preconditioning}

The conjugate gradient (CG) method is a well known numerical method from 
\cite{hestenes1952methods} for solving  systems of linear equations like \eqref{eq:const-skew-upd}. The matrix $\mathcal{B}$ as shown in \eqref{eq:large-mat} is symmetric and positive definite as shown in the previous section. This system is of the form $\mathcal{B} \operatorname{vec} (\mathbf{A}) = b$ where $b$ is the right-hand-side of the system shown in \eqref{eq:const-skew-upd}. The calculation of the expression, $\mathcal{B} \operatorname{vec} (\mathbf{A})$, can be accomplished without having to form a large $np \times np$ matrix as:
\begin{equation*}
\mathcal{B}\operatorname{vec} \left(\mathbf{A}\right) =  \operatorname{vec} \left(\sum_{g = 1}^G \!N_g \!\left( \bar{a}_g \!- \!\frac{1}{\bar{b}_g} \right) \left(\boldsymbol{\Sigma}_{g}^*\right)^{\!\!-1}\mathbf{A}\left( \boldsymbol{\Psi}_{\!g}^*\right)^{\!\!-1}\right).
\end{equation*}
Iterations of CG will usually require less operations in the form of $\mathcal{B}\operatorname{vec} \left(\mathbf{A}\right)$ shown above. Conjugate gradient applied to \eqref{eq:const-skew-upd} will converge in at most $np$ iterations. In practice, it doesn't take this many iterations to reach convergence, but there is still room for computational improvement. A simple preconditioner for $\mathcal{B}$ is:
\begin{equation}
P:=\sum_{g = 1}^G N_g \left( \bar{a}_g - \frac{1}{\bar{b}_g} \right) \left[ \boldsymbol{\Psi}_g^{-1} \otimes  \boldsymbol{\Sigma}_g^{-1}\right].
\end{equation}
Intuitively, $P$ is similar to the Jacobi preconditioner like what's discussed in \cite{barrett1994templates} except it is cheaper to form because $\boldsymbol{\Sigma}_g$ and $\boldsymbol{\Psi}_g$ are diagonal. In principle, the eigenvalues of $P$ are close to the eigenvalues of $\mathcal{B}$. As a result, the transformed system $P^{-1} \mathcal{B} \operatorname{vec} (\mathbf{A}) = P^{-1} b$ should have better conditioning and converge in less iterations of CG. Although not shown in this paper, preliminary tests suggest using this preconditioner decreases the number of iterations needed for convergence. Also, the value $\mathbf{A}$ in the previous iteration of the AECM algorithm can be used to initialize CG, meaning even less iterations are needed as the model approaches convergence, in principle.

\section{Simulation Studies}

\subsection{Simulation Study 1}

The primary goal of this study is to observe effective model recovery through the use of the BIC on a grid search through $(G, r, q) \in (1, 2, 3, 4, 5)$. A secondary goal of this study is to observe the accuracy of parameter estimation with the use of the AECM algorithm as sample size is increased.

\subsection*{Simulated Populations}

For this study, we generate 50 samples of size $N \in \{500, 1000\}$ from 12 different populations parameterized by a PMSBFA model in the skew-$t$ family. To facilitate the objective of this study, all simulated populations studied consist of 3 different groups or sub-populations in the proportions: $\boldsymbol{\pi} = (0.4, 0.4, 0.2)$. The common constrained models among these simulated populations were chosen to have the forms $\boldsymbol{C} = \{C_1, C_2, C_3, C_4\}$ where
\begin{equation*}
\begin{aligned}
C_1 = \mathrm{CCC\text{-}CCC\text{-}C\text{-}C},\quad&\quad
C_2 = \mathrm{CCC\text{-}CCC\text{-}U\text{-}C} \\
C_3 = \mathrm{CUC\text{-}CUC\text{-}U\text{-}C},\quad&\quad
C_4 = \mathrm{UCU\text{-}UCC\text{-}U\text{-}U}.
\end{aligned}
\end{equation*}
The common dimensions are from the set $\boldsymbol{d} = \{(10, 10), (10, 20), (30, 30)\}$. So, the 24 total simulated populations selected to be studied exhaust the cartesian product: $\boldsymbol{C} \times \boldsymbol{d}$.

The 12 populations consists of the following location parameters for components $g = 1,2,3$:
$$
\mathbf{M}_1 := \boldsymbol{0}, \quad
\mathbf{M}_2 := \boldsymbol{1}_n\boldsymbol{1}_p^\prime \quad
\mathbf{M}_3 := 3  \mathbf{M}_2,
$$
where $\boldsymbol{1}_n$ represents an $n \times 1$ vector with all entries equal to 1. 

For populations parameterized with constrained model $C_1$, the skewness matrices are constrained as $\mathbf{A}_g = \mathbf{A} = \boldsymbol{1}_n\boldsymbol{1}_p^\prime$. These sub-populations aren't well separated since the shared skewness causes the groups to skew into each other.

The skewness matrices for the other populations parameterized with models: $C_2$, $C_3$, and $C_4$ have skewness matrices that are defined as:
$$
\mathbf{A}_1 := -\boldsymbol{1}_n\boldsymbol{1}_p^\prime, \quad
\mathbf{A}_2 := \boldsymbol{U}_{n \times p}, \quad
\mathbf{A}_3 := \boldsymbol{1}_n\boldsymbol{1}_p^\prime,
$$
where $\boldsymbol{U}_{n \times p} := [u_{ij}]_{n \times p}$ where $u_{ij}$ is a random draw from a Uniform(-1, 1) distribution. The covariance and concentration parameters dependent upon the constrained model. These populations of matrix-variate data consist of mostly well separated populations so in order to put more of an emphasis on testing model recovery.

\subsection*{Covariance Matrices}

All 24 simulated populations were constructed with 2 row factors and 2 column factors. The populations that were constructed with constrained models $C_1$ or $C_2$ have the following parameters covariance parameters:
$\sigma = 1$, $\psi = 1$, $\boldsymbol{\Lambda} = \boldsymbol{U}_{n \times 2}$, and $\boldsymbol{\Delta} = \boldsymbol{U}_{p \times 2}$ (where $\boldsymbol{U}_{n \times k}$ represents a $n \times k$ matrix of realized draws from a Uniform(-1, 1)). So, for $g = 1, 2, 3$, we have that:
$$
\boldsymbol{\Sigma}_g = \sigma \boldsymbol{I}_n, \quad 
\boldsymbol{\Psi}_g = \psi \boldsymbol{I}_p, \quad
\boldsymbol{\Lambda}_g = \boldsymbol{\Lambda}, \quad
\boldsymbol{\Delta}_g = \boldsymbol{\Delta}, \quad
\nu_g = 5.
$$

The populations that were constructed with constrained models $C_3$ have the following parameters:
$(\sigma_1, \sigma_2, \sigma_3) = (1/4, \, 1,\, 4)$, $(\psi_1, \psi_2, \psi_3) = (4,\, 1, \, 1/4)$, $\boldsymbol{\Lambda} = \boldsymbol{U}_{n \times 2}$, and $\boldsymbol{\Delta} = \boldsymbol{U}_{p \times 2}$. So, for $g = 1, 2, 3$, we have that:
$$
\boldsymbol{\Sigma}_g = \sigma_g \boldsymbol{I}_n, \quad 
\boldsymbol{\Psi}_g = \psi_g \boldsymbol{I}_p, \quad
\boldsymbol{\Lambda}_g = \boldsymbol{\Lambda}, \quad
\boldsymbol{\Delta}_g = \boldsymbol{\Delta}, \quad
\nu_g = 5.
$$
The populations that were constructed with constrained models $C_4$ have the following parameters:
$\Sigma^{(n \times n)} = \operatorname{diag}(1, \, 4, \ldots, \, 1,\, 4)$, $\psi = 1$, and $(\nu_1, \nu_2, \nu_3) = (5, \, 10, \, 20)$. So, for $g = 1, 2, 3$, we have that:
$$
\boldsymbol{\Sigma}_g = \sigma_g \boldsymbol{I}_n, \quad 
\boldsymbol{\Psi}_g = \psi_g \boldsymbol{I}_p, \quad
\boldsymbol{\Lambda}_g = \boldsymbol{U}^g_{n \times 2}, \quad
\boldsymbol{\Delta}_g = \boldsymbol{U}^g_{p \times 2}, \quad
\nu_g = \nu_g.
$$
Where $\boldsymbol{U}^g_{n \times k}$ represents a $n \times k$ matrix of realized values from a Uniform(-1, 1), unique for the $g$th component.

\subsection*{Parameter Estimation}

For a given sample from a given population, the AECM algorithm is employed for all given combination of components, row factors, and column factors for 1 to 5 for all 256 possible variations of the constraints. The simulation results return the model that generates the best BIC. Table \ref{tab:simulation1} lists out their corresponding population dimension and constrained model. The tables list the total number of the correctly fitted components, row factors, and column factors from each of the best fit models generated by each of the 50 samples for a given sample size. The total number of true model selected (TMS) is also reported, and a form of a error estimate is reported for parameters corresponding to the 3 components within each given simulated population. The following norm $\|A\|_{\max} = \max_{i,j} |a_{ij}|$ is measured and the average is reported among the 50 random samples.

\begin{table}[t!]
\centering
\small
\label{tab:simulation1}
\renewcommand{\arraystretch}{1.35}
\setlength{\tabcolsep}{3pt}
\caption{Model recovery and estimation error summaries using the adaptive algorithm and BIC for mixtures of skew-$t$ distributions. Each entry reports true model selection frequency (TMS), correct recovery of $G$, average adjusted Rand index (ARI), and the average over 50 datasets of the maximum element-wise estimation errors for $\mathbf{M}_g$ and $\boldsymbol{\Psi}_g^* \otimes \boldsymbol{\Sigma}_g^*$ among $g = 1, 2, 3$.}
\begin{tabular}{>{\centering\arraybackslash}m{1.2cm} *{6}{>{\centering\arraybackslash}m{2.15cm}}}
\toprule
 & \multicolumn{2}{c}{\textbf{10 $\times$ 10}} & \multicolumn{2}{c}{\textbf{10 $\times$ 20}} & \multicolumn{2}{c}{\textbf{30 $\times$ 30}} \\
\cmidrule(lr){2-3}\cmidrule(lr){4-5}\cmidrule(lr){6-7}
 & \textbf{$N=500$} & \textbf{$N=1000$} & \textbf{$N=500$} & \textbf{$N=1000$} & \textbf{$N=500$} & \textbf{$N=1000$} \\
\midrule
\rotatebox{90}{\footnotesize\textbf{CCC-CCC-C-C}} & \makecell[l]{TMS: 50 \\ G: 50 \\ ARI: 0.874 \\ M: 0.525 \\ $\Psi\otimes\Sigma$: 0.177} & \makecell[l]{TMS: 49 \\ G: 50 \\ ARI: 0.870 \\ M: 0.384 \\ $\Psi\otimes\Sigma$: 0.096} & \makecell[l]{TMS: 50 \\ G: 50 \\ ARI: 0.931 \\ M: 0.604 \\ $\Psi\otimes\Sigma$: 0.325} & \makecell[l]{TMS: 49 \\ G: 50 \\ ARI: 0.934 \\ M: 0.426 \\ $\Psi\otimes\Sigma$: 0.203} & \makecell[l]{TMS: 47 \\ G: 48 \\ ARI: 0.982 \\ M: 0.652 \\ $\Psi\otimes\Sigma$: 0.733} & \makecell[l]{TMS: 47 \\ G: 49 \\ ARI: 0.984 \\ M: 0.465 \\ $\Psi\otimes\Sigma$: 0.462}
\\[1.2ex]
\rotatebox{90}{\footnotesize\textbf{CCC-CCC-U-C}} & \makecell[l]{TMS: 50 \\ G: 50 \\ ARI: 1.000 \\ M: 0.786 \\ $\Psi\otimes\Sigma$: 0.156} & \makecell[l]{TMS: 50 \\ G: 50 \\ ARI: 1.000 \\ M: 0.530 \\ $\Psi\otimes\Sigma$: 0.083} & \makecell[l]{TMS: 50 \\ G: 50 \\ ARI: 1.000 \\ M: 0.887 \\ $\Psi\otimes\Sigma$: 0.285} & \makecell[l]{TMS: 50 \\ G: 50 \\ ARI: 1.000 \\ M: 0.584 \\ $\Psi\otimes\Sigma$: 0.174} & \makecell[l]{TMS: 50 \\ G: 50 \\ ARI: 1.000 \\ M: 0.896 \\ $\Psi\otimes\Sigma$: 0.610} & \makecell[l]{TMS: 50 \\ G: 50 \\ ARI: 1.000 \\ M: 0.657 \\ $\Psi\otimes\Sigma$: 0.379}
\\[1.2ex]
\rotatebox{90}{\footnotesize\textbf{CUC-CUC-U-C}} & \makecell[l]{TMS: 50 \\ G: 50 \\ ARI: 1.000 \\ M: 0.994 \\ $\Psi\otimes\Sigma$: 0.617} & \makecell[l]{TMS: 50 \\ G: 50 \\ ARI: 1.000 \\ M: 0.665 \\ $\Psi\otimes\Sigma$: 0.353} & \makecell[l]{TMS: 50 \\ G: 50 \\ ARI: 1.000 \\ M: 1.142 \\ $\Psi\otimes\Sigma$: 1.113} & \makecell[l]{TMS: 50 \\ G: 50 \\ ARI: 1.000 \\ M: 0.763 \\ $\Psi\otimes\Sigma$: 0.650} & \makecell[l]{TMS: 50 \\ G: 50 \\ ARI: 1.000 \\ M: 1.137 \\ $\Psi\otimes\Sigma$: 1.692} & \makecell[l]{TMS: 50 \\ G: 50 \\ ARI: 1.000 \\ M: 0.820 \\ $\Psi\otimes\Sigma$: 1.075}
\\[1.2ex]
\rotatebox{90}{\footnotesize\textbf{UCU-UCC-U-U}} & \makecell[l]{TMS: 50 \\ G: 50 \\ ARI: 1.000 \\ M: 2.206 \\ $\Psi\otimes\Sigma$: 1.587} & \makecell[l]{TMS: 50 \\ G: 50 \\ ARI: 1.000 \\ M: 1.528 \\ $\Psi\otimes\Sigma$: 0.862} & \makecell[l]{TMS: 50 \\ G: 50 \\ ARI: 1.000 \\ M: 2.328 \\ $\Psi\otimes\Sigma$: 1.673} & \makecell[l]{TMS: 50 \\ G: 50 \\ ARI: 1.000 \\ M: 1.696 \\ $\Psi\otimes\Sigma$: 0.894} & \makecell[l]{TMS: 50 \\ G: 50 \\ ARI: 1.000 \\ M: 2.991 \\ $\Psi\otimes\Sigma$: 2.699} & \makecell[l]{TMS: 50 \\ G: 50 \\ ARI: 1.000 \\ M: 2.147 \\ $\Psi\otimes\Sigma$: 1.503}
\\[1.2ex]
\bottomrule
\end{tabular}
\end{table}

\section{Real Data Analysis}

\subsection{MNIST data set}

The MNIST data set is a collection of handwritten digits represented by $28 \times 28$ pixel matrices with light intensities ranging from 0 to 255. The original data set includes 60,000 pixel matrices as training data and another 10,000 as testing data. Typically, this full data is used as an example dataset for "black box" models such as neural networks, tree models, SVMs, etc. This paper will utilize various subsets of the images from the training data set to fit models discussed in this paper. First, consider an example from \cite{gallaughermcnicholas2020} in which 25 samples of 400 digits, 200 images of each of the digits 1 and 2. These 25 samples of digits were constructed so they have no overlap, and to facilitate clustering for the models tested in this paper, the light values that make up the pixel matrices of the images are scaled to be between 0 and 10. However, a good number of the cells are 0 for all of the pixel images within any given sample. So, to avoid either $\boldsymbol{\Sigma}_g^*$ or $\boldsymbol{\Psi}_g^*$ from becoming singular, a layer of standard normal white noise is added onto the pixel matrices in order to support parameter estimation and facilitate good clustering. 


For these initial 25 samples, parameter estimation was performed for $G = 2$ and the adaptive approach to model selection was performed for all possible combinations of row and column factors from 1 to 20 for both PMSBFA and PMMVBFA models. Additionally, on the same 25 samples of digits, the cases for 25 percent and 50 percent classification were also considered. Table \ref{tab:mnist12_ari} reports the summary statistics in regards to the models' classification success. At all tested levels of supervision, the PMSBFA models outperform the PMMVBFA models on each of the 25 samples considered.

\begin{table}[ht]
\centering
\caption{The average Adjusted Rand Index (ARI) and the average Misclassification rate across 25 samples of 400 of the MNIST digits 1 and 2 for varying levels of supervision.}
\label{tab:mnist12_ari}
\begin{tabular}{lc|c}
\hline
Model & PMSBFA & PMMVBFA \\
\hline
Supervision & ARI (MCR) & ARI (MCR) \\
\hline
0\% (clustering) & 0.846 (0.040) & 0.713 (0.078) \\
25\% & 0.905 (0.024) & 0.863 (0.036) \\
50\% & 0.938 (0.016) & 0.897 (0.026) \\
\hline
\end{tabular}
\end{table}

\subsection{Olivetti Face Set}

We lastly consider the Olivetti Face dataset, consisting of greyscale images of faces that were taken between 1992 and 1994 at AT\&T laboratories in Cambridge. The images come from 40 individuals with 10 images of each person with a resolution of 64 $\times$ 64 pixels. There are a total of 400 images with varied lighting and facial expressions, and some of photos also depict faces with glasses. We fit all PMSBFA models for $G = 1, \ldots, 5$ and all row and columns factors between 25 to 40. The BIC chose a model with 2 components, 37 row factors, and 35 column factors. Additionally, the constraints $\boldsymbol{\Delta}_g = \boldsymbol{\Delta}$ and $\mathbf{A}_g = \mathbf{A}$ were chosen (I.e. the model $\mathrm{UUU}$-$\mathrm{CUU}$-$\mathrm{C}$-$\mathrm{U}$). The components have the mixing proportions $\pi_1 = 0.7025$ and $\pi_2 = 0.2975$. Heatmaps of the mean matrices of each component are provided which also indicate the key distinctions between the two subpopulations.

\begin{figure}[ht]
    \centering
    \includegraphics[width=0.8\textwidth]{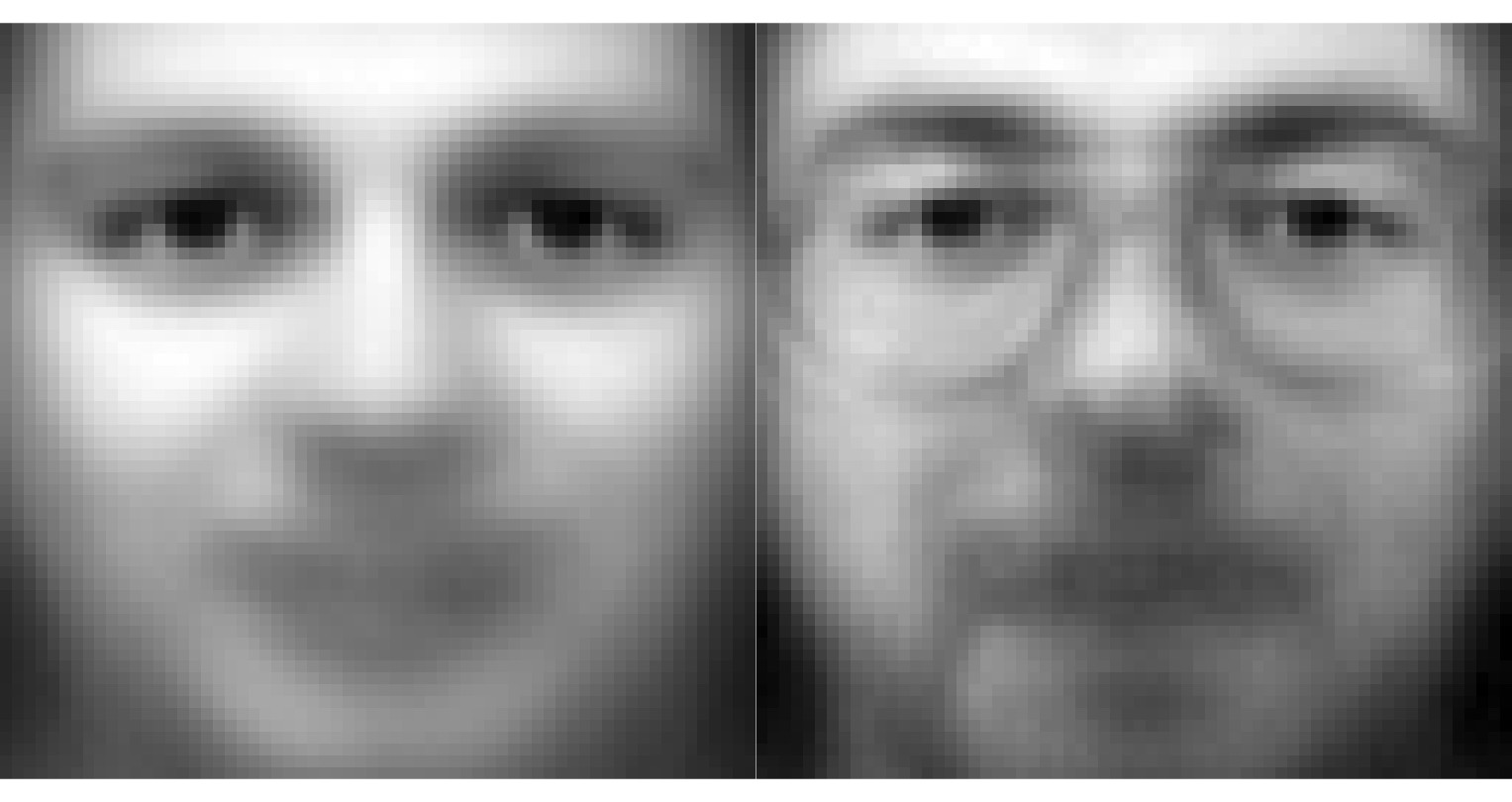}
    \caption{Depictions of the estimated mean fields for the PMSBFA model selected by the BIC. On the left is the heatmap of the mean for component 1, and on the right is a heatmap of component 2.}
    \label{fig:oliv}
\end{figure}

The signal of glasses appears prominently in component 2, and the selected PMSBFA model clustered the images based on whether or not the face was wearing glasses. There were only two misclassifications based on this criterion. The non-parsimonious models from \cite{gallaughermcnicholas2019} result in the BIC choosing a model with just one component which doesn't provide much information. Therefore, this example shows why there is a need for further dimension reduction through parsimony.

\section{Discussion}

The PMSBFA family, comprised of 256 models, has been introduced with skew $t$ distribution. In the simulated settings that were studied, the models introduced exhibited excellent model recovery and parameter estimation that becomes more accurate as the sample size increased. In real data analysis, PMSBFA models outperform PMMVBFA models in terms of classification. Additionally, the Olivetti face data set illustrates why PMSBFA models are a necessary extension of the skewed mixtures from \cite{gallaughermcnicholas2019}. As the sizes of the model families grow, one important future consideration is a search algorithm that can reduce the number of unnecessary model fits without sacrificing model selection accuracy.


\bibliographystyle{apalike} 
\bibliography{Mix_Fac}  

\end{document}